\documentclass{iopart}
\usepackage{graphicx}
\begin{document}
\title[Classical wave experiments on chaotic scattering]
{Classical wave experiments on  chaotic scattering}

\author{U. Kuhl\dag, H.-J. St\"ockmann\dag, R. Weaver\ddag}

\address{\dag Fachbereich Physik der Philipps-Universit\"at Marburg,
Renthof 5, D-35032 Marburg, Germany}

\address{\ddag Department of Theoretical and Applied Mechanics,
University of Illinois, 104 S. Wright Street, Urbana, Illinois 61801, USA}

\begin{abstract}
We review recent research on the transport properties of classical
waves through chaotic systems with special emphasis on microwaves
and sound waves. Inasmuch as these experiments use antennas or
transducers to couple waves into or out of the systems, scattering
theory has to be applied for a quantitative interpretation of the
measurements. Most experiments concentrate on tests of predictions
from random matrix theory and the random plane wave approximation.
In all studied examples a quantitative agreement between experiment
and theory is achieved. To this end it is necessary, however, to
take absorption and imperfect coupling into account, concepts that
were ignored in most previous theoretical investigations.
Classical phase space signatures of scattering are being examined in
a small number of experiments.
\end{abstract}


\submitto{\JPA}

\ead{ulrich.kuhl@physik.uni-marburg.de}
\ead{r-weaver@uiuc.edu}

\maketitle

\section{Introduction}

Experiments with classical waves, originally conceived to study
the spectral properties of closed systems, have more recently become
an important tool to study the scattering properties of open
systems as well.

At about 1980 the main experimental information on spectra and
scattering properties of chaotic systems came from nuclear physics
and resulted in the development of random matrix theory by Wigner,
Dyson, Mehta and others on one side \cite{por65}, and of
scattering theory by Weidenm\"uller  and coworkers on the other
side \cite{mah69}.

Interest in the properties of chaotic systems suddenly
renewed, when it was conjectured by Casati \etal \cite{cas80}
and Bohigas \etal \cite{boh84b} that the universal properties of
all chaotic systems should be described by random matrix theory.
An equivalent of the conjecture for open systems was formulated
somewhat later by Bl\"{u}mel and Smilansky \cite{blue90b}. These
conjectures proved to be extremely useful and gave a new impetus
to the whole field.

The experimental situation, however, remained unsatisfactory until
1990, when the first analogue experiments were published, starting
with vibrating solids \cite{wea89a} and chaotic microwave cavities
\cite{stoe90}. Meanwhile the technique has been extended to water
surface and pressure waves \cite{blue92a, chi96} and optical
systems \cite{doy02a,din02,gen05}. At about the same time the
tremendous progress in the fabrication of semiconductor
microstructures allowed to study the transport properties of
submicron size structures of various types such as quantum dots
\cite{mar92}, tunnelling barriers \cite{fro94}, and quantum
corrals \cite{cro95}. The state of the art till 1999 can be found
in chapter 2 of reference \cite{stoe99}.

In the beginning these experiments focussed on the spectral
properties of closed systems though there was already an early
experiment looking for the scattering properties of an open
microwave system \cite{dor90}. Strictly speaking, the systems are
always open, due to the necessity to introduce antennas or
transducers to perform the measurements. Therefore scattering
theory is mandatory for a quantitative interpretation of the
measurements. This is why it became necessary to establish a
relation between the Green's function of the closed system and the
scattering matrix \cite{ste95}, which is a direct equivalent of a
corresponding expression derived in nuclear physics many years
ago. For the case of an isolated resonance the well-known
Breit-Wigner formula is recovered \cite{alt93b}. Details will be
given in section 2. The same equivalence also holds for
quantum-dot systems, as long as electron-electron interactions can
be neglected. Thus with respect to scattering theory it does not
make any difference whether atomic nuclei, quantum dots, microwave
cavities, or vibrating blocks are involved, provided one is
interested in the universal features only.

This allows experimental tests of theoretical predictions from
scattering theory with an unprecedented precision. A number of
important consequences of scattering theory have been demonstrated
with classical waves, but never found in nuclei, such as the
algebraic decay of the scattering matrix from a system with a
small number of channels \cite{alt95a, lob03a}, or the fact that
with increasing coupling to the channels the resonances eventually
become narrow again \cite{per00}, known as resonance trapping.

There are a number of clear advantages of microwave resonators and
vibrating blocks as compared e.\,g.\ to nuclei, but also to quantum
dots. First of all, the geometry is perfectly known, and the
coupling to the external channels can be calculated explicitly
from the antenna geometry. The systems are clean, and if there are
impurities, they are introduced intentionally. The geometry can
easily be varied thus allowing level dynamic measurements of
different types, which is impossible for nuclei, and can be
achieved only in a very restricted way in quantum dots. Further
advantages arise from the fact that the relevant wavelengths in
microwave resonators or vibrating blocks are of the order of mm to
cm resulting in very convenient system sizes. Last but not least
the experiments can be performed at room temperature since there
are no dephasing processes as in quantum dots which have to be
frozen out. This advantage is lost in microwave experiments using
superconducting cavities, but still the applied temperatures of
some K are far above the 0.01 to 1 K temperatures usually
applied in quantum dot experiments.

In this review we present results from microwave resonators,
vibrating blocks, and optical systems,
and their comparison with the predictions of scattering theory.
Wave experiments on localization in disordered
systems are excluded since they are treated in a separate
contribution to this volume \cite{gen05b}. Results on quantum dots,
tunneling diodes and other mesoscopic structures are excluded as
well. They would have deserved a contribution of their own, and
will be mentioned only in cases where a comparison with results
from experiments with classical waves seems appropriate.

This review is organized as follows. In the next section a Green's
function technique is applied to establish a relation between the
scattering matrix and the Green's function of the system. In
addition the random superposition of plane waves model is
introduced to describe amplitude distributions in chaotic
cavities. Though demonstrated for the example of a flat microwave
cavity where there is a perfect analogy to quantum mechanics, the
results of this section can be applied without changes to all
experiments with classical waves. Therefore this section is the
basis for all following ones where the results in open microwave
cavities, vibrating blocks, and optical systems are presented.

\section{The scattering matrix for billiard systems}
\label{sec:SMatrix4Bils}
\subsection{The billiard Breit-Wigner formula}
\label{subsec:BreitWigner}

A microwave billiard is a special example for a scattering system,
where the measuring antennas take  the role of the scattering
channels. We are now going to discuss the consequences of the
external coupling for resonance positions and widths.

Let us start with a two-dimensional quantum billiard of arbitrary shape, and
assume that there are a number of channels coupled to the billiard
with diameters small compared to the wave length. Let
$\psi(\bi{r},k)$ be the amplitude of the field within the billiard
while matter waves enter and exit through the different channels.
$\psi(\bi{r},k)$ obeys the Helmholtz equation
\begin{equation}\label{eq:2.1}
\left(\Delta +k^2\right)\psi=0\,.
\end{equation}
Close to the channels the field is given by a superposition of
incoming and outgoing circular waves,
\begin{equation}\label{eq:2.2}
\psi(\bi{r},k)=a_iH_0^{(1)}(k_i|\bi{r}-\bi{r_i}|)
-b_iH_0^{(2)}(k_i|\bi{r}-\bi{r_i}|)\,,
\end{equation}
where $\bi{r_i}$ is the position of the $i$th channel. $a_i$
and $b_i$ are the amplitudes of the waves entering and leaving the
billiard through the channel, and $H_0^{(1)}(x)$ and
$H_0^{(2)}(x)$ are Hankel functions. Denoting the amplitude
vectors of the waves entering or leaving the billiard by
$a=(a_1,a_2,\dots,a_N)$ and $b=(b_1,b_2,\dots,b_N)$, respectively,
the scattering matrix for the billiard system is  defined by the
relation
\begin{equation}\label{eq:2.4a}
b=Sa\,.
\end{equation}
The $a_i$ and $b_i$ are obtained by matching fields and their
normal derivatives at the coupling positions of the channels.
This is achieved by means of Green's function techniques.
Details can be found, e.\,g., in Section~6.1.2 of Ref.~\cite{stoe99}.
Thus one obtains
\begin{equation}\label{eq:2.11}
S=\frac{1-\imath W^\dag GW}{1+\imath  W^\dag GW}\,,
\end{equation}
where the matrix elements $W_{nm}$ of $W$ contain the information
of the coupling of the $n$th eigenfunction to the $m$th channel,
and $G$ is given by
\begin{equation}\label{eq:2.12}
    G=\frac{1}{k^2-H}\,,
\end{equation}
where $H=-\Delta$ is the Hamilton operator for the undisturbed
system.

An explicit calculation of the $W_{nm}$ has to take into account the
details of the coupling geometries. For antennas in microwave
resonators this has been achieved in Ref. \cite{bar05a}. But it is
common practice to treat the $W_{nm}$ as free parameters which are
taken from the experiment.

An elementary transformation of equation~(\ref{eq:2.11}) yields the
equivalent expression
\begin{equation}\label{eq:2.13}
    S=1-2iW^\dag\frac{1}{k^2-H+\imath WW^\dag} W\,,
\end{equation}
For the case of non-overlapping resonances and point-like coupling
this reduces to
\begin{equation}\label{eq:2.15}
S_{ij}(k)=\delta_{ij}-2\imath\gamma\bar{G}(\bi{r_i},\bi{r_j},k)\,,
\end{equation}
with the modified Green's function
\begin{equation}\label{eq:2.16}
\bar{G}(\bi{r_i},\bi{r_j},k)=\sum_n\frac{\psi_n(\bi{r_i})\psi_n(\bi{r_j})}
{k^2-k_n^2+\imath\gamma\sum_k\left|\psi_n(\bi{r_k})\right|^2}\,
\end{equation}
(see reference \cite{ste95}).
$\psi_n$ are the real eigenfunctions of the closed systems
and $\gamma$ describes the antenna coupling. The last equations
constitute the billiard equivalent of the Breit-Wigner formula
well-known from nuclear physics for many years~\cite{bla52}. Often
the scattering matrix is studied on the time domain. To this end
the Fourier transform is performed,
\begin{equation}\label{eq:2.17}
\hat{S}_{ij}(t) = \frac{1}{2\pi}\int S_{ij} (k)\rme^{-\imath\omega
t}\,\rmd\omega \,,
\end{equation}
where $\omega=\hbar k^2/2m$. It follows
\begin{equation}\label{eq:2.18}
\hat{S}_{ij}(t)
=\delta_{ij}\,\delta(t)-\frac{\hbar\gamma}{m}
   \sum_n\psi_n(\bi{r_i})\psi_n(\bi{r_j})
\exp{\left[-\left(\imath \omega_n+\frac{1}{2}\Gamma_n\right)t\right]}\,,
\end{equation}
with $\Gamma_n=(\hbar\gamma/m)\sum_k\left|\psi_n(\bi{r_k})\right|^2$.
This holds for the quantum mechanical case.
For non-dispersive classical waves we have instead $\omega=ck$,
and $\gamma$ has to be replaced by $k \gamma$ in
Eqs.~(\ref{eq:2.15}) and (\ref{eq:2.16}). Thus one obtains
\begin{equation}\label{eq:2.19}
\hat{S}_{ij}(t) =\delta_{ij}\,\delta(t)-
2\gamma c \sum_n \psi_n(\bi{r_i})\psi_n(\bi{r_j})
\sin\left(\omega_n t\right)\rme^{-\frac{1}{2}\Gamma_n t}\,,
\end{equation}
where now $\Gamma_n=\gamma c \sum_k\left|\psi_n(\bi{r_k})\right|^2$.
In both cases $\Gamma$ corresponds to the decay rate of the square of the field.

Equation (\ref{eq:2.15}) shows that the complete modified Green's
function $\bar{G}(\bi{r_i},\bi{r_j},k)$ can be obtained from a
transmission measurement between two antennas of variable
position~\cite{ste95}. A reflection measurement at one antenna only
yields the spectrum and the modulus of the wave
function~\cite{stoe90,ste92}. The `true' wave function
$G(\bi{r_i},\bi{r_j},k)$, however, is never obtained since the
resonances are broadened by $\Gamma_n$.
This is the price to be paid for every measurement.

\subsection{The random plane wave approximation}
\label{subsec:RPWA}

A model which proved to be very useful to describe field
distributions and correlation in chaotic billiards is the  random
plane wave approximation. It assumes that the wave function can be
described at any point, not too close to the boundary, by a random
superposition of plane waves, entering from different directions,
but with the same modulus of the wave number,
\begin{equation}\label{eq:2.20}
    \psi(\bi{r})=\sum\limits_na(\bi{k_n})\rme^{\imath \bi{k_n}\bi{r}}\,,\qquad
    |\bi{k_n}|=k\,.
\end{equation}
The model has been independently introduced in different fields,
among others in acoustics \cite{ebe84} and in  quantum mechanics
of chaotic systems \cite{ber77a}. It is known as Berry's
conjecture in the quantum chaos community. The random plane wave
approximation can be obtained from the Green's function of the
billiard by averaging over a small energy window as was shown by
Hortikar, Srednicki \cite{hor98a} and further exploited by Urbina,
Richter \cite{urb03}.

As an immediate consequence of the central limit theorem one obtains
a Gaussian distribution for the amplitudes of the wave function of
closed systems,
\begin{equation}\label{eq:2.40}
    p(\psi)=\sqrt{\frac{A}{2\pi}}\exp\left(-\frac{A\psi^2}{2}\right)\,,
\end{equation}
where $A$ is the billiard area (see e.\,g. Section 6.2 of
Ref.~\cite{stoe99}). In open systems real and imaginary part of
$\psi(r)$ are uncorrelated and both Gaussian distributed.

The model allows the explicit calculation of various correlation
functions, the most popular among them the spatial amplitude
autocorrelation function
\begin{eqnarray}\label{eq:2.41}
    c_\psi(\bi{r_1},\bi{r_2})&=&
    \langle\psi^*(\bi{r_1})\psi(\bi{r_2})\rangle\nonumber\\
    &=&\frac{1}{A}J_0(kr)\,,
\end{eqnarray}
where $r=\left|\bi{r_1}-\bi{r_2}\right|$.

The random plane wave approximation can be used to calculate the
distribution of line widths. The line width in a billiard with $\nu
$ attached channels is given by a sum over the squares of $\nu$
Gaussian distributed random numbers, (compare the expressions
following equations~(\ref{eq:2.18}) and (\ref{eq:2.19})). The
distribution $p_\nu(x)$ is hence a $\chi^2$ distribution
\begin{equation}\label{eq:2.42}
p_\nu(x)=\left(\frac{A}{2}\right)^{\nu/2} \frac{1}{\Gamma\left(\nu/2
\right)}x^{\nu/2-1} \rme^{-\frac{A}{2}x}\,.
\end{equation}

The distribution of resonance depths for a reflection measurement
can be obtained as well.
According to equation~(\ref{eq:2.16})
the resonance depths are proportional to
$\left|\psi_n(\bi{r})\right|^2$, where $\bi{r}$ is the antenna
position. The depth distribution function for a closed system is
thus a $\chi^2$ distribution with $\nu=1$,
\begin{equation}\label{eq:2.43}
p_1(x)=\sqrt{\frac{A}{2\pi x}}\rme^{-\frac{A}{2}x}\,,
\end{equation}
i.\,e. a Porter-Thomas distribution. For an completely open system one
obtains a $\chi^2$ depth distribution with $\nu=2$, which is
just a single exponential,
\begin{equation}\label{eq:2.45}
    p_2(x)=\frac{A}{2}\rme^{-\frac{A}{2}x}\,.
\end{equation}

The model allows the calculation of averages of transients of
various correlations of scattering matrix elements as well. From
equation~(\ref{eq:2.18}) we obtain for the average of
$|\hat{S}_{ij}(t)|^2$
\begin{equation}\label{eq:2.46}
    \hat{C}_{ij}(t)=\langle |\hat{S}_{ij}(t)|^2 \rangle
    \sim\left(1+\frac{4\gamma t}{\hbar A}\right)^{-(2+\nu/2)}\,,
\end{equation}
i.\,e. an algebraic decay is expected with an exponent reflecting
the number of attached channels. For a very large number of
channels the $\chi^2$ distribution concentrates in a narrow peak,
which is again a consequence of the central limit theorem. In this
limit there is a cross-over in the decay of $\hat{C}_{ij}(t)$ to a
single exponential behaviour, as is immediately evident from
equation~(\ref{eq:2.18}),
\begin{equation}\label{eq:2.47}
    \hat{C}_{ij}(t)
    \sim \rme^{-\Gamma t}\,.
\end{equation}
It follows that the Fourier transform of $\hat{C}_{ij}(t)$, which
due the the convolution theorem of Fourier theory is just
the energy autocorrelation function of the respective scattering
matrix elements, is described by a Lorentzian. This is the regime
of the Ericson fluctuations \cite{eri63}, well-known in nuclear
physics for many years.
It is noteworthy that the same universal fluctuation statistics
were long ago noted in acoustics as well \cite{sch62a,sch62b}.

\subsection{Acoustic and elastodynamic systems}
\label{subsec:acoustic}

The expressions (\ref{eq:2.1}) to (\ref{eq:2.47}) of the previous subsections
have direct analogs in acoustic and elastodynamic systems.
The concepts of ray propagation, and the
more abstract formalisms, are essentially identical. The differences are
chiefly in the governing wave equation~(\ref{eq:2.1}) and in the details of
experimental transduction.

Like microwaves in 2-d billiards, waves in 2-d shallow water tanks
\cite{rol85,ebe84} are governed by equation~(\ref{eq:2.1}), with
$k^{2}$ interpreted as $\omega ^{2}/gh$ ($g$: acceleration of
gravity, $h$: water depth). Chinnery \cite{chi96} presents acoustic
measurements in a deep water tank in quasi-2-d cylindrical
geometries. These experiments differ from the corresponding 2-d
microwave experiments in that they generally have non-Dirichlet
boundary conditions at tank edges and lower $Q$'s than do microwaves
in billiards. A similar wave equation applies to the acoustics of a
drum head, for which Teitsworth \cite{tei00} has compared random
matrix theory (RMT) predictions, for level statistics and
wavefunctions. Capillary waves \cite{blue92a} have even lower $Q$'s,
and an interpretation of $k^{2}$ as $(\omega ^{2}\rho /T)^{1/3}$,
where $T$ is the surface tension. There is at least one report of
wave chaos in a soap bubble \cite{arc98}, for which the governing
equation is again (\ref{eq:2.1}), with $k^{2}$ interpreted as
$\omega ^{2} T / \rho h$, plus corrections for air loading.

There are surprisingly few reports of wave chaos in 3-d acoustic
systems. One example is seismic waves, i.\,e., low frequency elastic
waves in the earth, are also governed by equation
(\ref{eq:weaver1}). At short enough wavelength they are scattered by
disorder and show mesoscopic behaviors. Further there are
experiments in gasses, e.\,g., dry helium or argon, which are worth
exploring, as impedance mismatches with boundaries can be made large
and $Q$'s correspondingly large also. It should be mentioned that
there is a large literature on acoustics in rooms
(e.\,g.~\cite{kut91,ber92b,ebe82}), however very little of it is
carried out in a context of RMT. Notable exceptions are the works of
Schr\"oder \cite{sch54,sch59,sch69a,sch69b}, Lyon \cite{lyo69}, and
Davy \cite{dav90}.

The chief acoustic venue for experimental studies of wave chaos and
related mesoscopic wave physics has been elastodynamics, i.\,e.,
ultrasonics in  solids. At MHz frequencies wavelengths are of the
order of several mm in typical solids, and laboratory systems are of
convenient size. In aluminum, $Q$'s can approach $10^{5}$,
\cite{wea89double,del94,ell95,and01,lob03a}, in quartz they approach
$10^{6}$ \cite{ell96,ber99b}. The governing equation differs from
(\ref{eq:2.1}) in that the dependent variable is a not a scalar, but
rather vector displacement $\bi{u}$, and $\Delta$ is replaced by a
tensor-valued linear differential operator, $\pounds_{ij}$
\cite{gra75}

\begin{eqnarray}
\label{eq:weaver1} \omega^{2} \rho (\bi{x})
u_{i}(\bi{x},\omega ) - \pounds_{ij} u_{j}(\bi{x},\omega )
= -\rho (\bi{x})f_{i}(\bi{x},\omega) \\\nonumber
\pounds_{ij}=-\partial _{k} c_{ikjl} (\bi{x}) \partial _{l}
\end{eqnarray}
where $\bi{f}$ is the density of externally applied force
divided by the mass density $\rho$, and summation convention is in force.
In isotropic media $\pounds$ is given in terms of
the Lam\'{e} moduli of elasticity $\lambda$ and $\mu$,

\begin{equation}\label{eq:weaver1a}
\pounds_{ij} = -(\lambda  \delta _{ik} \delta _{jl} +\mu \delta _{ij
} \delta _{kl} +\mu  \delta _{il} \delta _{jk})
\partial _{k} \partial _{l}.
\end{equation}
Propagation and reflection of rays differs
from the simpler case of equation~(\ref{eq:2.1}) in that there are two
propagation speeds corresponding the transverse and longitudinal
polarizations (more in anisotropic media) and reflection generally
results in mode conversion. A ray of one type incident on a wall
will reflect into two or more types. Thus ray dynamics is complex.

In 2-d, e.\,g., in a plate below the first cutoff frequency, and
in the presence of up/down symmetry in material properties and in
geometry the governing equations decouple into two classes of
waves. One may be identified as flexural; it has a displacement
that is predominantly normal to the surface. It has a governing
equation -- at sufficiently long wavelength - like (\ref{eq:2.1})
with $\Delta$ replaced by $\Delta^2$, the Kirchhoff plate
equation. These flexural waves propagate with high dispersion.
Except for the dispersion and the modified boundary conditions,
such systems closely resemble the familiar scalar billiard. They
have been discussed in the context of ray chaos and semiclassics
and RMT by Bogolmolny and Hugues \cite{bog98}. The other decoupled
wave has displacement vector predominantly in-plane, and consists
of superposed shear and dispersive-longitudinal motions, coupled
by mode conversion at the boundaries, the Poisson plate equation.
At frequencies above the first cutoff ($f = c_{T}
/2\times\mathrm{thickness}$) the elastodynamics of a plate is best
described in terms of its several dispersive Lamb modes of
propagation \cite{gra75}, all coupled at the boundaries.

Regardless of the precise form of the governing equation or the
geometry, and indeed regardless of the presence of internal
heterogeneities, all these systems may be abstracted in the limit of
no absorption and in the presence of the proper boundary conditions,
by a self-adjoint operator problem. With typical, e.\,g.,
traction-free, boundary conditions, the governing equation
(\ref{eq:weaver1}) remains not only Hermitian, but real and
symmetric, i.\,e., time-reversal invariant, and the eigenvalues
$\omega _{n}^{2}$ are real and non negative, and the eigenfunctions
$\bi{u}^{n}(\bi{x})$ real and orthogonal.

\begin{eqnarray}
\label{eq:weaver2} H_0  \vert \bi{u}^{n}\rangle
= \omega _{n}^{2} \vert \bi{u}^{n} \rangle;
\qquad H_0 = (1/\rho )\pounds\\ \nonumber
\langle \bi{u}^{n}  \vert \bi{u}^{m} \rangle=
\int \bi{u}^{n}(\bi{x}) \bi{\cdot} \bi{u}^{m}(\bi{x})\rho
(\bi{x}) \rmd\bi{x} = \delta _{nm}
\end{eqnarray}
The mass density $\rho$ acts as a weight function in the inner product.
The Hamiltonian $H_0$ is real-symmetric positive semi-definite.
It is conjectured for many purposes to be ergodically equivalent
to a member of the Gaussian Orthogonal Ensemble of random matrices.

Linear loss mechanisms are modeled by visco-elasticity, a general
framework based upon a presumption that stress depends only on the
history of local strain. In this case the otherwise Hermitian
operator $H_0$ gains an additional term, an imaginary symmetric
operator $-\rmi WW^T$, which may in principle depend on frequency.
Such dependence is, however, expected to be weak. On expressing
$WW^T$ in terms of its (real) eigenvectors $\vert
\bi{c}\rangle$ and eigenvalues $\chi_c$, one writes

\begin{eqnarray}
\label{eq:weaver3} H = H_0 - \rmi WW^T = H_0 - \rmi \Sigma _{c}  \vert
\bi{c}\rangle  \chi_{c} \langle \bi{c}\vert
\end{eqnarray}

This may be compared to the effective Hamiltonian in the
denominator of (\ref{eq:2.13}). The eigenvectors $\bi{c}$
represent loss channels, the real quantities $\chi_c$ their
strengths. The channels $|\bi{c}\rangle$ may be represented as
normalised vector functions of position $\bi{c}(\bi{x})$.
The $\chi_c$ and $|\bi{c}\rangle$ are expected to vary slowly
with frequency. Some of the loss channels may be associated with
physical contacts to the body, supports or attached sensors.
Others may be internal and material. The non-Hermitian but
symmetric operator $H$ has complex orthogonal eigensolutions
\begin{eqnarray}
\label{eq:weaver4} H  \vert \bi{u}^{n} \rangle = (\omega
_{n}-\rmi\Gamma _{n}/2)^{2} \vert \bi{u}^{n} \rangle \\ \nonumber
\langle \bi{u}^{n} \vert \bi{u}^{m} \rangle=\int
 \bi{u}^{n}(\bi{x}) \bi{\cdot} \bi{u}^{m}(\bi{x})
 \rho(\bi{x}) \rmd\bi{x} = \delta _{nm}
\end{eqnarray}
It may be noted that the
above inner product is defined without complex conjugation.

Level widths are given exactly by (compare the expression
following equation~(\ref{eq:2.19})),
\begin{equation}\label{eq:weaver5}
\Gamma _{n} = \frac{1}{\omega_{n}} \sum_{c}
\frac{\chi_{c} \vert \langle \bi{u}^{n}  \vert \bi{c}\rangle\vert ^{2}}
{\langle  \bi{u}^{n\ast } \vert \bi{u}^{n} \rangle}
\end{equation}
There is a similar identity for the natural frequencies
\begin{equation}\label{eq:weaverno2}
\omega _{n}^{2}-\frac{\Gamma _{n}^{2}}{4} =
\frac{\vert \langle \bi{u}^{n\ast } \vert H_0  \vert \bi{u}^{n} \rangle \vert}
{\langle \bi{u}^{n\ast }  \vert \bi{u}^{n} \rangle}
\end{equation}
Green's function $(\omega ^{2} - H )^{-1}$ is given by a modal
expansion, cf equation~(\ref{eq:2.16}),
\begin{equation}
\label{eq:weaver6} G=\sum\limits_n {\frac{\vert
\bi{u}^n\rangle\langle \bi{u}^n\vert }{\omega ^2-(\omega
_n -\rmi\Gamma _n /2)^2}}
\end{equation}
or
\begin{equation}
\label{eq:weaver7} \tilde {G}(x,y;\omega )=\sum\limits_n
{\frac{\bi{u}^n(x)\bi{u}^n(y)}{\omega ^2-(\omega _n
-\rmi\Gamma _n /2)^2}}
\end{equation}
In the time domain one has,
\begin{eqnarray}
\label{eq:weaver8} G(x,y;t)=-{\rm Im}\sum\limits_n
{\frac{\bi{u}^n(x)\bi{u}^n(y)\exp \{-\rmi\omega _n t-\Gamma
_n t/2\}}{\omega _n -\rmi\Gamma _n /2}} \\\nonumber G_{ij}
(x,y;t)=-{\rm Im}\sum\limits_n {\frac{\bi{u}_i^n
(x)\bi{u}_j^n (y)\exp \{-\rmi\omega _n t-\Gamma _n t/2\}}{\omega
_n -\rmi\Gamma _n /2}}
\end{eqnarray}

Elastic waves are coupled into and out of the bodies by, most often,
piezoelectric devices. The ideal device exerts a concentrated point impulse
in response to an electrical impulse. In practice the dynamical force is
distributed in time and over the face of the transducer. Similarly an ideal
detector responds with an electrical signal proportional to the
instantaneous displacement at a point. In practice a detector gives a signal
dependent on the recent history of displacement across the transducer face.
Calculating the space-time transfer function is difficult. It is a
complicated problem in piezoelectricity and elastodynamics \cite{kin87}. It
depends on how the transducer is coupled and on the material of the body.
Nevertheless, general arguments using electro-mechanical reciprocity show
that a detector $\bi{d}$ connected to a single coaxial cable may be represented as
delivering a signal
\begin{equation}
\label{eq:weaver9} V = \chi_{d} \langle \bi{d} \vert \bi{u}\rangle
\end{equation}
linear in the displacement field. The detection channel $\langle
\bi{d} \vert$ may be presented as a vector function of position,
non zero only at the position of the physical detector and oriented in
the direction of its sensitivity. Similarly a source $\bi{s}$
delivers a distributed force (normalised to unit mass density)
$\bi{f}(\bi{x})$ in response to an applied voltage $V$, which may be
expressed as
\begin{equation}
\label{eq:weaver10} \vert \bi{f} \rangle=\vert \bi{s}\rangle  \chi _{s} V
\end{equation}
where $\vert \bi{s}\rangle$ represents the spatial distribution and
orientation of the force $\bi{f}(\bi{x})$ generated by the source
$V$. $| \bi{s}\rangle$ and $| \bi{d}\rangle$ as well as the
scalars $\chi_d$ and $\chi_s$
in general depend on frequency, but do so weakly on the scale of the
level spacings and over the ranges of fluctuation expected in the
wave response of the structures.

Unlike the usual case with microwaves, our sources and detectors
do not present significant compliance to the specimen.
What impedance there is, is almost entirely mechanical and
independent of the electronics. Signal measurements \textit{per
se} therefore do not absorb or affect the waves. While the contact
devices may so affect the waves, it is usually found that the
vectors $\vert \bi{d}\rangle$ and $\vert \bi{s}\rangle$ do not correspond
to significant loss channels $\vert \bi{c}\rangle$. Thus unlike the microwave
case we do not speak of measuring $S$ matrices but rather of
responses. The signal in a detector $\bi{d}$ due to an input
$V^{\rm in}$ to a source $\bi{s}$ is the solution $V^{\rm out}$ to
\begin{equation}
\label{eq:weaver11} H| \bi{u}\rangle-\omega ^{2}  \vert \bi{u}\rangle
=\vert \bi{f}\rangle
= \vert \bi{s}\rangle  \chi _{s} V^{in}
\end{equation}
\begin{equation} \label{eq:weaverno3}
V^{\rm out}=\chi_{d} \langle \bi{d} | \bi{u} \rangle
\end{equation}
i.\,e.,
\begin{equation}\label{eq:weaver12}
V^{\rm out}=
\chi_{d} \langle \bi{d} \vert G\vert \bi{s}\rangle  \chi _{s} V^{\rm in }
\end{equation}
The output signal is the input signal convolved with the $ds$ component
of the Green's function, convolved with the source and receiver
functions $\chi_s$ and $\chi_d$, respectively. It is for this reason
that we speak of measuring components of the Green's function, not
components of an S matrix. In acoustics literature the process is
not usually termed scattering or measurement of an S matrix element,
although it is scattering in the sense of this review article.

The approximation of negligible loss into the channels $\bi{d}$ and $\bi{s}$ is
exact if source and detector are non-contact.   Laser interferometry
provides a non contact, albeit less sensitive detection \cite{ros00}.
Dropped ball bearings \cite{and01,wea89double,del94}, and laser thermoelastic
excitation are also sources which do not act as loss channels.
The use of contact pin-like piezoelectric transducers
\cite{wea89double,del94,ros00,dra97}, or needles \cite{sch03b,and01} is only
minimally intrusive.

The random plane wave approximation (\ref{eq:2.20}) does not apply directly to
elastodynamics, as there is additional information needed on the relative
amplitudes of the different wave types. Nevertheless, RMT tells us (for the
non lossy case where the modes are real) that a diffuse wave field in
general, and a mode in particular, should be a Gaussian random process. The
generalized Berry conjecture \cite{ako04a} asserts
that the two-point mode correlation function is the imaginary part of the
Green's function
\begin{equation}
\label{eq:weaverno4}
-\frac{1}{\pi}{\rm Im}G_{ij} (x,y\,,\omega _n)
=
\frac{\partial N/\partial \omega}{2\omega} \, \overline {\bi{u}_i^n (x)\,\bi{u}_j^n (y)} \,,
\end{equation}
where $\partial N/\partial \omega$ is the spectral density of modes.

This is equivalent to equation (\ref{eq:2.41}), but applies also to
non-scalar waves and to regions near heterogeneities. That this is
exact if the average is over a frequency interval, i.\,e., for short
times, was pointed out in \cite{ako04a}.

In the presence of losses, the modes are in general not real. But the above
equation tells us that the real and imaginary parts are uncorrelated. That
the imaginary parts are Gaussian processes also has been assumed \cite{lob00b},
but the relative variance of the real and imaginary parts is as yet not
understood.

As in the discussion preceding equation~(\ref{eq:2.42}) and
(\ref{eq:weaver5}), one notes that a level width is given by a
weighted sum of the squares of Gaussian random numbers. If all
channels have equal strength $\chi$, and if the modes are
approximately real (as they are in the weak absorption limit) then
a line width $\Gamma$ is a chi square random number, and $\vert
V^{\rm out}(t)\vert ^{2}$ decays in time like an incoherent
superposition of terms, each $\sim \exp(-\Gamma _{n}t)$
\begin{equation}
\label{eq:weaver13} \overline {\left| {V^{\rm out}(t)} \right|^2}
=\int {p(\Gamma)\rmd\Gamma \exp (-\Gamma t)} =E_o
\left(1+2\overline \Gamma t/\nu \right)^{-\nu /2}
\end{equation}
where $\nu$ is the number of channels and $\overline \Gamma$ is the
average decay rate. This was first derived in acoustics by Schr\"oder \cite{sch65}
and later by Burkhardt \cite{bur96,bur97,bur98}. This differs from (\ref{eq:2.46}) due to the assumed
lack, in acoustics, of significant losses into the source and detection
channels. As in (\ref{eq:2.47}), it reduces to a simple exponential in the limit of a
large number of channels $\nu$.

In section \ref{sec:SoundWaves} we review the many experimental evidences in support of the
applicability of RMT, for open (lossy) and closed (nonlossy) acoustic
systems. Many of the earliest studies emphasized eigenstatistics, in
particular level and eigenfunction statistics. These are reviewed for
completeness. More recent research has emphasized \textit{responses}, i.\,e., scattering.
Response statistics are often measured with the intention to understand them
in terms of eigenstatistics. Thus the two types of studies are closely
related.

\section{Microwave experiments}\label{sec:MicrowaveExperiments}

\subsection{Fundamental tests of scattering theory}\label{subsec:tests}

There are a number of fundamental predictions from scattering
theory on resonance shapes, width distributions, decay behaviour,
and others, all of which have been verified in microwave systems.
We give here only a short account, as this subject has already
thoroughly been discussed in section 6 of reference \cite{stoe99}.

Alt \etal \cite{alt96a} performed a detailed study of the shapes of
resonance in the non-overlapping regime and could show that they are
described perfectly well by the billiard Breit-Wigner function
(\ref{eq:2.15}). In a quarter-stadium-shaped microwave billiard with
three attached antennas the same group looked for the distribution
of the line depths at one antenna, and the sum of the line depths at
two and three antennas, respectively, and found $\chi^2$
distributions with $\nu =1,2,3$ as expected \cite{sch95b,ric99}. In
the time domain one expects for the system an algebraic decay with a
power of 7/2 due to the presence of three antennas, see
equation~(\ref{eq:2.46}), which was also observed in \cite{alt95a}.
Algebraic decay behaviour has been observed in reverberant blocks as
well (see Section~\ref{subsec:SoundDissipation}).

The first microwave scattering experiment has been performed by
Doron \etal \cite{dor90} in an elbow-shaped resonator. It will be
discussed in more detail in section \ref{subsec:class} in the proper
context. It is interesting to note, that essentially the same
system was studied numerically independently by Jalabert \etal
\cite{jal90} but at this time in the context of universal
conductance fluctuations, a nice illustration of the fact that
nuclei, microwave billiards, and mesoscopic systems do not differ,
as long as universal scattering properties are concerned.

A measurement of the elements of the scattering matrix yield
directly the billiard Green's function, though modified due to the
presence of the antennas (see equation~(\ref{eq:2.15})). Thus the
complete quantum mechanical information is available. This was used
in reference \cite{ste95} in a measurement in a quarter stadium
billiard, where one antenna was fixed, whereas the other one was
scanned through the billiard. From the transmission
spectra the propagator is obtained by means of Fourier transforms.
For short times a circular wave was found, emitted from the fixed
antenna, being destroyed after a small number of reflections. After
some time the primary pulse was partially reconstructed due to the
focussing properties of the circular part of the boundary. Similar
measurements have been performed in a ray-splitting billiard,
realised by a rectangular billiard, where a quarter-circular insert
of teflon was attached to one corner \cite{sch01c}. Teflon has an
index of refraction of $n=1.44$, thus giving rise to a ray splitting
upon every reflection. In microwave billiards the matrix elements
$W_{nm}$ of $W$, describing the coupling of the antennas to the wave
function, see equation~(\ref{eq:2.11}), can be explicitly calculated
from the geometry \cite{stoe02c,bar05a}. Since the poles of the scattering
matrix $S$ are given by the eigenvalues of the effective Hamiltonian
$H_{\rm eff}=H-\imath WW^\dag$, see equation (\ref{eq:2.13}), this
allows a detailed study of these poles in dependence of the coupling
strengths \cite{per00}. Details will be given in section
\ref{subsec:pole}.

The $\chi^2$ distribution found for the line widths presents an
indirect verification of the random plane wave approximation. A nice
direct verification, though in the regime of visible light, has been
obtained by Doya \etal \cite{doy02a}. The authors studied the
transport of a He-Ne laser beam through a glass fiber with a
$D$-shaped cross-section with $R=63.5\,\mu m$. For the near-field
intensity directly at the exit of the glass fiber they found a
speckle-like pattern. Introducing an additional lens, and mapping
the intensity pattern in the focal plane, the far-field intensity
was obtained. It corresponds to the modulus squared of the
two-dimensional Fourier transform of the near-field amplitude.
The far-field intensity was found to
concentrate on a circle with no specific preference for any
direction, thus demonstrating the validity of the random plane wave
approximation.

\subsection{Pole distributions in the complex plane}
\label{subsec:pole}

In the case of  resonance overlap the Breit-Wigner formula is not
valid anymore (see equation~(\ref{eq:2.16})), and one has to
consider the full $S$-matrix as given in equation~(\ref{eq:2.12})
resulting in new and partly counterintuitive behaviour such as the
resonance trapping effect.


\begin{figure}
\includegraphics[width=\columnwidth]{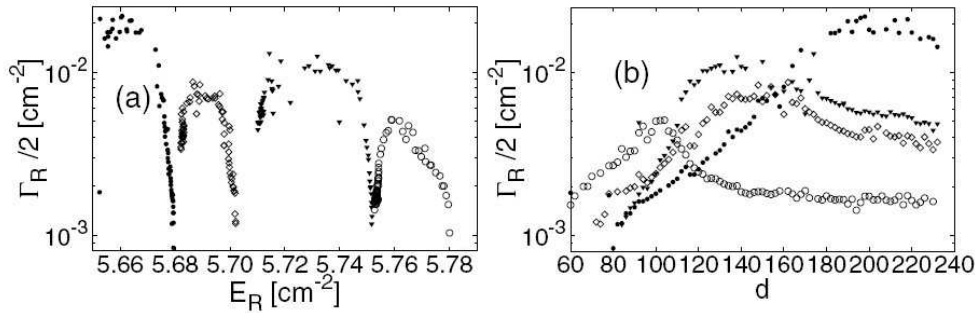}
\caption{\label{fig:ResonanceTrapping} Resonance position for four
resonances in a Sinai billiard with attached waveguide with variable
opening (a) and their corresponding width $\Gamma_R/2$ versus the
slit width $d$ (b). The different resonances are marked by different
symbols (taken from reference \cite{per00}). }
\end{figure}

Different scales of lifetimes appear for the resonant states if there
is sufficient overlap \cite{mol67,des99,jun99b}. Some of the states
align with the decay channels and become short lived, while the
remaining ones decouple from the continuum and become long lived, a
phenomenon called trapping.
The doorway states in nuclear physics provide an example for the
alignment of the short-lived states with the channels
\cite{sok97,per99b}. For a clear experimental demonstration of the
trapping effect, the coupling strength to the decay channels has to
be tuned. This was done experimentally by Persson \etal
\cite{per00}, a detailed theoretical discussion can be found in
\cite{stoe02c}. In the experiment the microwaves were coupled to a
resonator of the shape of a Sinai billiard via an attached
waveguide, and the reflection was measured. Using a slit with
variable opening width $d$ the coupling from the waveguide to the
billiard could be changed. The slit width was varied in steps of
0.1\,mm from 60 to 240\,mm. Resonance positions and widths were
extracted using a centered time-delay analysis \cite{per00}. In
figure~\ref{fig:ResonanceTrapping}(a) the dependence of four
resonance positions on the opening $d$ of the slit is shown. In
figure~\ref{fig:ResonanceTrapping}(b) the width $\Gamma$ is plotted
as a function of the slit opening $d$. For small slit openings $d$
the widths of the resonances first increase with $d$ but for larger
openings the width finally decrease again, thus demonstrating the
effect of resonance trapping \cite{per00}.


In the case of resonance trapping all resonances are treated on
equal footings. If there is only an overlap of two resonances and
all others resonances are far apart one can describe the occurring
pattern in the transmission by Fano resonances \cite{lee99,mag03}.
Fano resonances have been observed originally in photoabsorption
in atoms \cite{beu35,fan35,fan61} but occur in many fields in
physics like electron and neutron scattering \cite{ada49,sim63},
Raman scattering \cite{cer73}, photoabsorption in quantum well
structures \cite{fai97a}, scanning tunnel microscopy \cite{mad98},
and ballistic transport through quantum dots ("artificial atoms")
\cite{goer00,noec94,rot03b,kob02}. Close to a given resonance
$E_i^R$ the spectra can be approximated by the Fano form
\cite{fan61,kim01c},
\begin{equation}
\label{eq:TFano} |T (E, d)|^2  \approx \frac{|E -
E_i^R(d) + q_i (d) \Gamma_i (d) /2|^2}{\left[ E - E_i^R (d)
\right]^2 + \left[ \Gamma_i (d)/2 \right]^2}
\end{equation}
where $d$ describes the coupling strength, $E_i^R (d)$ is the
position of the $i$th resonance, $\Gamma_i (d)$ its width, and $q_i
(d)$ the complex Fano asymmetry parameter.
Window resonances appear in the limit $q \rightarrow 0$ while the
Breit-Wigner limit is reached for $|q| \gg 1$ (see reference
\cite{rot04a} for details). It should be noted that, in general, $q$
cannot be simply identified with the ratio of resonant to
non-resonant coupling strength \cite{kim01c,kim03d,eic03}. Moreover,
since Fano resonances result from the interference between
resonances related to the eigenmodes in the cavity, the parameter
$q$ depends very sensitively on the specific constellation of the
involved resonance poles \cite{lee99,mag03}.

In none of the previously mentioned experiments the Fano resonances
could be systematically studied  by varying the  coupling strength.
Again this has been achieved in a microwave billiard
\cite{rot04a,rot}. The setup consisted of two commercially available
waveguides which were attached centred both to the entrance and the
exit side of a rectangular resonator. At the junctions to the cavity
metallic apertures of different openings $d$ were inserted. In the
applied frequency range only two transverse modes for the
rectangular cavity could be excited. A continuous transition from a
narrow Breit-Wigner ($q$ very large, $d$ small) to a window
resonance ($q \to 0$, $d$ large) was found \cite{rot04a}. In a
further investigation the effect of decoherence and dissipation on
the Fano resonance lineshapes was discussed \cite{rot}.


Up to now only resonances with no or weak overlap or, as in the case
of the Fano resonances, with just two strongly overlapping
resonances have been investigated experimentally. For a complete
exploration of the features of the scattering matrix an
investigation of the poles in the strong overlapping regime is
needed. By means of the harmonic inversion technique
\cite{wal95,man97a,man97b,mai99} this now has become possible
\cite{kuha}.

\begin{figure}
\includegraphics[width=\columnwidth]{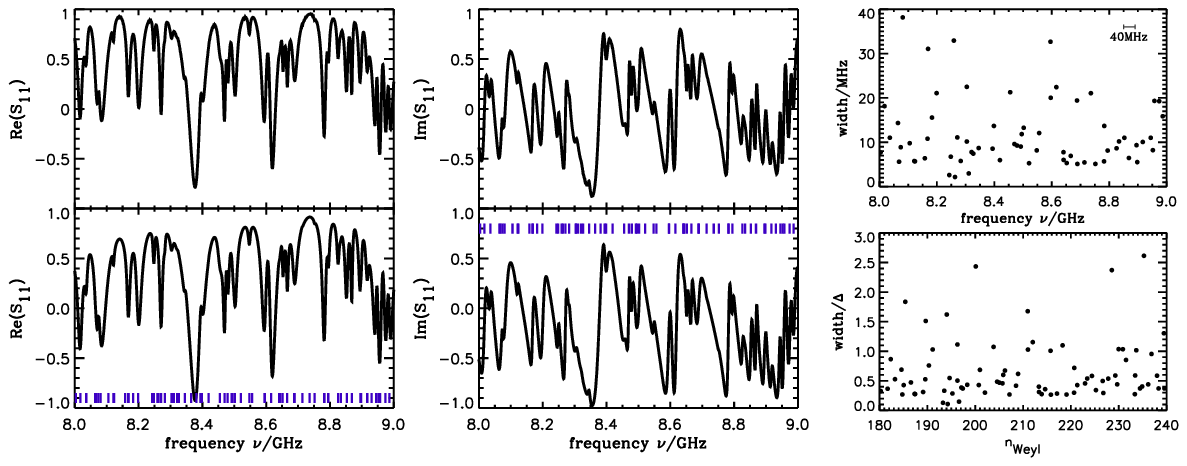}
\caption{\label{fig:ResonancePoles} Left and centre: Real and
imaginary part of a spectrum of a half Sinai microwave billiard with
absorption. On top the measured spectrum is shown, on bottom the
reconstructed spectrum as obtained from the harmonic inversion. The
resonance positions are marked by short lines. Right: Resonance
widths for the displayed spectra in MHz (top), and normalised to the
mean level spacing (bottom) (taken from reference \cite{kuha}). }
\end{figure}

In figure~\ref{fig:ResonancePoles}(left and centre) the complex
reflection coefficient $S_{11}$ of a half Sinai billiard is shown
for the frequency range from 8 to 9 GHz (for details see \cite{kuha}).
It is compared with a reconstructed spectrum obtained
from the harmonic inversion. The resonance positions are marked by
small lines. Apart from a secular overall phase shift the
reconstruction is nearly perfect. Global phase shifts are not
serious as long as one is only interested in resonance positions,
width and absolute values of the amplitude, and their behaviour in
dependence of an external parameter. On the right (top) the widths
are plotted versus the frequency and (bottom)  versus energy,
normalised to a mean level spacings of one. In the example the
resonance widths are of the order of the mean level spacing
$\Delta$, but it was possible to penetrate into regimes where the
resonance width amounted up to about ten times the mean level
spacings. For isolated resonances one expects a Porter-Thomas
distribution \cite{por56,por65}, whereas for strong coupling
the tails of the distribution decay according to $P_\Gamma(\Gamma)
\propto \Gamma^{-2}$ \cite{fyo97b,som99}. This could be verified in
the experiment \cite{kuha}.

\subsection{Distribution of reflection and transmission coefficients}
\label{subsec:PSPT}

The first experimental investigation of scattering theory was done
by Doron \etal who studied the reflection properties of an elbow
system  \cite{dor90}. It concentrated  on the influence of the
classical escape behaviour on the reflection, and will be treated
in subsection \ref{subsec:class} in the proper context.
It took about ten more years until the influence of coupling and
absorption was investigated experimentally by several groups in
microwave billiards in more detail
\cite{men03a,kuh05,bar05a,bar05b,hem05a,hem05b}. Barth\'elemy \etal
developed a model to describe the coupling of a microwave antenna
to the system and could separate their contribution $\Gamma_c$
to the total resonance width $\Gamma$ from the part $\Gamma_{ohm}$
resulting from ohmic losses in the wall \cite{bar05a}.

\begin{figure}
\includegraphics[width=\columnwidth]{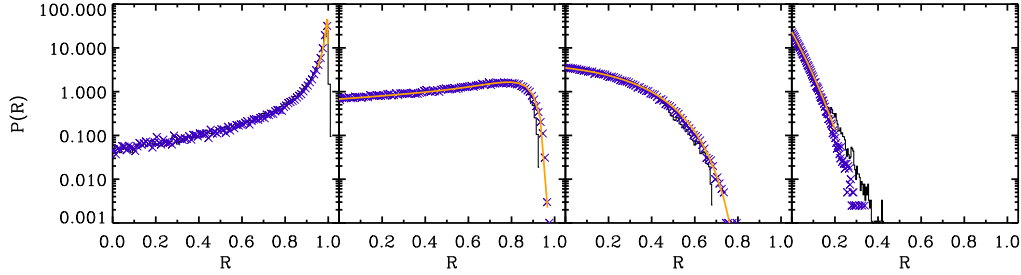}
\caption{\label{fig:PRTheta}
The distribution of the reflection $P_R(R)$
is shown for four different coupling strength ($T_a$=0.116, 0.754, 0.989, 0.998)
and total absorption ($\gamma$=0.56, 2.42, 8.40, 48).
Additionally the theoretical curve (yellow line) and
numerical results are plotted (blue crosses)
(for details see \cite{men03a,kuh05}).}
\end{figure}

M\'ende{z-S\'a}nchez \etal investigated reflection spectra in
dependence of the coupling strength and the absorption
\cite{men03a}. The experiment was performed in a half Sinai
billiard, where an additional absorbing wall could be introduced
to increase the total absorption. They could verify the
theoretical predictions for the limiting case of strong absorption
\cite{kog00}. The known result of weak absorption \cite{bee01b}
could not be verified as it assumed perfect coupling which was not
realised in the experiment. An interpolating
formula for the distribution $P_{R,0}$ of reflection coefficients
in the case of perfect coupling was derived \cite{kuh05,fyo04}
\begin{equation}\label{eq:PR0}
P_{R,0}(R_0) =  C_\beta \frac{ \rme^{-{\alpha}/{(1-R)}}
}{(1-R)^{2+\beta/2}} \left[ A \alpha^{\beta/2-1} + B (1-R)^{\beta/2}
\right],
\end{equation}
with $\alpha={\gamma_d \beta}/{2}$, $A={\alpha}
\left(\rme^{\alpha}-1\right)$, $B=( 1 +\alpha - \rme^{\alpha} )$,
$C_{\beta} =[A \Gamma(1+\beta/2, \alpha)/\alpha^2 + B
\rme^{-\alpha}/\alpha]^{-1}$ and the universality index $\beta$.
$\gamma_d$ is a parameter describing the dissipation \cite{kuh05}
and $\Gamma(x,\alpha) =\int_\alpha^\infty t^{x-1}\rme^{-t}dt$ is the
upper incomplete Gamma function. Meanwhile there exists an exact
but lengthy expression by Savin, Sommers, and Fyodorov \cite{sav05}. The
deviations from equation~(\ref{eq:PR0}) are typically small. From the
distribution $P_{R,0}$ the joint distribution of the reflection
coefficient and the phase of the reflection coefficient
($S=\sqrt{r} \rme^{\rmi\theta}$) for incomplete coupling can be calculated via
\begin{equation}\label{eq:PRTheta}
P_{R,\Theta}(S) =  \left( \frac{1-\langle S \rangle^2}{|1-S \langle S \rangle|^2} \right)^2
           \frac{1}{2\pi} P_{R,0}(R_0),
\end{equation}
where $\langle S \rangle = \sqrt{1-T_a}$ and $R_0=|S_0|^2$ with
$S_0=(S-\langle S \rangle)/(1-\langle S \rangle S)$. In
figure~\ref{fig:PRTheta} the distribution of the reflection $P(R)$
is shown for four different coupling strength $T_a$ and absorption
$\gamma_d$.
The theoretical curves shown in
blue are obtained via a projection to the corresponding axis of
$P_{R,\Theta}(S)$. A good agreement between experiment and theory
is found for $P(R)$ \cite{men03a,kuh05}. In a further work the phase
distribution $P(\Theta)$, called the Poisson kernel, and the joint
distribution were determined experimentally, again in perfect
agreement with theory \cite{kuh05}.
For phase distributions see section~\ref{subsec:SoundComplexEigenmodes}.

\begin{figure}
\includegraphics[width=\columnwidth]{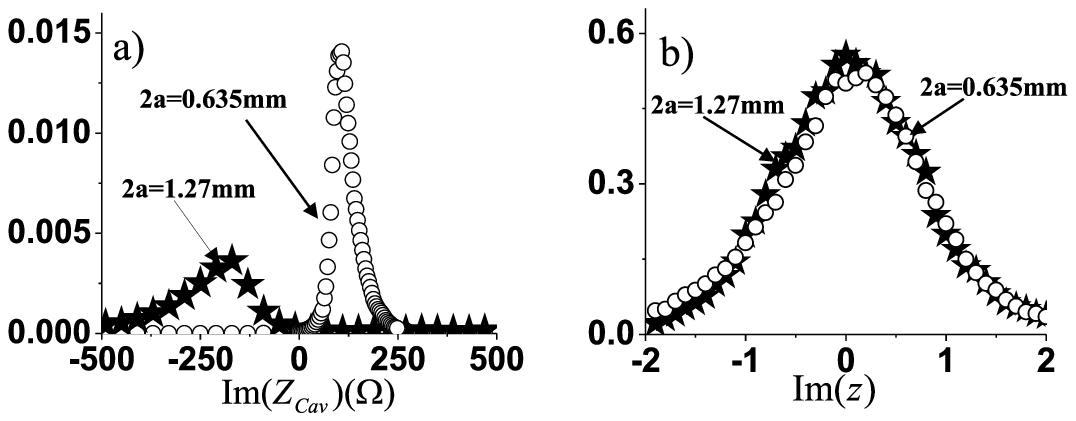}
\caption{\label{fig:PImpedance} (a) Distribution of the imaginary
part of cavity impedance for two different couplings from 9 to 9.6
GHz. (b) Rescaled plot of two curves in (a) after dividing the
experimental impedances by the impedances $Z_0$ of the antenna
(taken from reference \cite{hem05a}). }
\end{figure}

Hemmady \etal applied a different approach \cite{hem05a,hem05b,hem,zhe}.
They did not study the scattering matrix but looked for the impedance
matrix $Z$ instead. These quantities are related via
\begin{equation}\label{eq:ImpedanceMatrix}
Z=Z_0(1+S)/(1-S)\,.
\end{equation}
$Z_0$ describes the coupling port and is
fixed by an independent measurement. The impedance matrix relates
the applied voltages $V$ to the currents
at the antennas ($V=ZI$), whereas the scattering
matrix relates the incoming flux with the outgoing flux. Using
the impedance matrix Hemmady \etal could remove the effects of
coupling thus obtaining a normalised cavity impedance, which shows
a universal behaviour \cite{hem05a,hem}. This is demonstrated by
figure~\ref{fig:PImpedance}.

Microwave graphs are another possibility to study the influence of absorption
on the distribution of reflection coefficients, real and imaginary part
of impedances etc. This has been done by Hul \etal \cite{hula}.
Again a perfect agreement with RMT predictions was found.

Though there are no analytical results available for the joint
distribution of scattering matrix elements for more than one
channel, apart from the distribution of the modulus of the eigenvalues \cite{bro97a},
there are results for the distribution of the total
transmission $T$. This quantity is of practical interest, since
due to the Landauer formula the total transmission through an open
quantum dot is, up to an universal vector, nothing but the
conductance $G$,
\begin{equation}\label{trans01}
    G\sim T=\sum\limits_{n,m} \left|S_{nm}\right|^2\,,
\end{equation}
where the index $n$ sums over all incoming and the index $m$ over all
outgoing channels. There are a number of predictions
on the distribution $p(T)$ of the transmission in dependence of the
channel number for systems with and without time-reversal symmetry
\cite{bar94,jal94}. In the calculation the matrix elements had
been assumed to be Gaussian distributed, and uncorrelated, apart
from the constraint of the universality of the $S$ matrix. The
assumption of Gaussian distributed scattering matrix elements
seems to be in contrast to expression (\ref{eq:2.11}) for the
scattering matrix. It can be shown, however, that for chaotic
cavities and perfect coupling both approaches are equivalent
\cite{bro95b} (see section IIA.4 of reference \cite{bee97} for a
discussion). Experimentally obtained conductance distributions for
quantum dots were in serious disagreement with the calculations,
though the influence of the break of time reversal symmetry could
be qualitatively reproduced \cite{hui98a}. A possible problem
arises from the fact, that the geometry of the dots defined by the
gate electrodes is ill-defined, and  thus  the channel number as
well.

In microwave cavities this problem does not exist. Experiments
have been performed in a chaotic cavity with two attached
waveguides both on the entrance and the exit side. Time reversal
symmetry was broken by inserting ferrite cylinders. The
phase-breaking mechanism of the ferrite can be qualitatively
understood as follows: Applying a magnetic field, the electronic
spins in the ferrite perform a precession with the consequence
that microwaves reflected from a ferrite surface experience a
phase shift whose sign depends on the direction of propagation.
This effect has been used previously by So \etal \cite{so95} and
Stoffregen \etal \cite{sto95b} to study time-reversal symmetry
breaking in closed microwave billiards. However, the ferrite
introduces strong absorption, and the phase-breaking effect is
present only in a narrow frequency window in the wing of the
ferromagnetic resonance. Thus the calculation of the transmission
distribution has to be modified by introducing additional
absorbing channels. If this is done, both the break of
time-reversal symmetry and the channel number dependence can be
well reproduced by the calculations \cite{sch05a}.

\subsection{Correlations of $S$ matrix elements}\label{subsec:corr}

The Fourier transform of the spectral autocorrelation function

\begin{equation}\label{corr01}
C(E)=\left<\rho(\bar{E}) \rho(\bar{E}+E)\right> -\left(\left<\rho
(\bar{E})\right>\right)^2\,,
\end{equation}
yields the spectral form factor
\begin{equation}\label{corr02}
K(t)=\int_{-\infty}^\infty C(E)\exp\left(-\frac{\imath}{\hbar}
Et\right)\,dE\,.
\end{equation}
In chaotic systems $K(t)$ shows a hole for small values of $t$,
resulting from the spectral rigidity. Extensive work on the
spectral hole has been done, among others in nuclear physics
\cite{lom94} and in the analysis of spectra from microwave
cavities \cite{alt97b}. The spectral hole is thus a convenient
indicator to test chaotic behaviour.

In open systems the spectral autocorrelation function has to be
replaced by scattering matrix correlation functions

\begin{equation}\label{corr03}
C[S^*_{ab},S_{cd}](E) = \langle S^*_{ab}(\bar{E})\,
S_{cd}(\bar{E}+E)\rangle -
   \langle S^*_{ab}(\bar{E})\rangle\; \langle S_{cd}(\bar{E})\rangle \,,
\end{equation}
or, alternatively, by their Fourier transforms

\begin{equation}\label{corr04}
\hat{C}[S^*_{ab},S_{cd}](t) = \langle \hat{S}^*_{ab}(t)\,
\hat{S}_{cd}(t)\rangle -
   \delta(t)\langle S^*_{ab}(\bar{E})\rangle\; \langle S_{cd}(\bar{E})\rangle \,,
\end{equation}
where
\begin{equation}\label{corr05}
\hat{S}_{ab}(t)=\int_{-\infty}^\infty
S_{ab}(E)\exp\left(-\frac{\imath}{\hbar} Et\right)\,dE
\end{equation}\,.

The optical theorem
\begin{equation}\label{corr06}
    \sigma_{\rm tot}^{(a)}=2\left(1-{\rm Re} S_{aa}\right)
\end{equation}
establishes a relation between the diagonal elements of the
scattering matrix and the total cross section, often used in
nuclear physics. Correlation functions of the diagonal elements of
the scattering matrix may thus be alternatively expressed in terms
of correlation functions of the total cross section \cite{gor02a}.

\begin{figure}
\includegraphics[width=\textwidth]{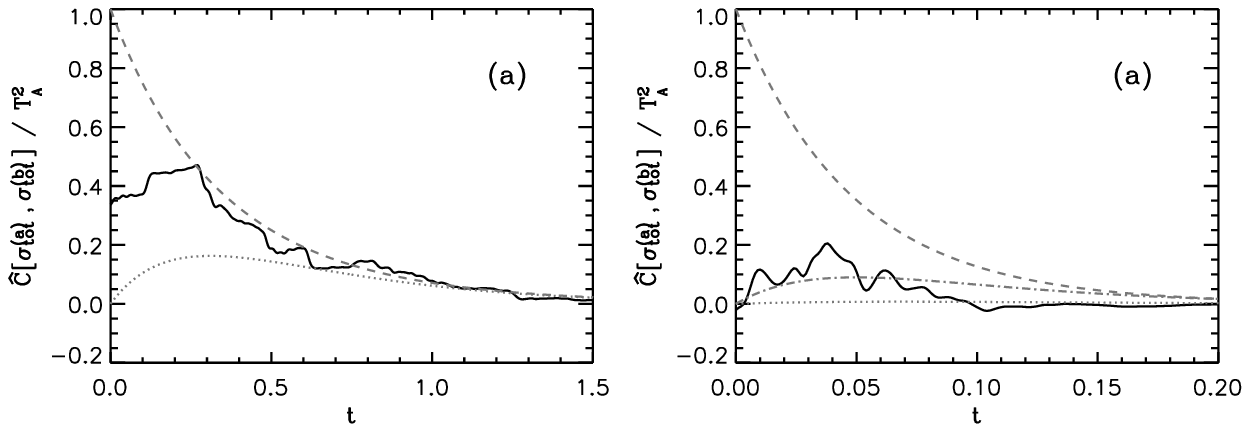}
\caption{\label{fig:corrhole}
Normalized cross-correlation function $\hat C[\sigma_{\rm tot}^{(a)},\sigma_{\rm tot}^{(b)}]$
for (a) a rectangular and (b) a chaotic billiard with $C_3$
symmetry with the Fourier transform taken over the frequency range
13 to 14\,GHz (the  division by $T_A^2$ accounts for
the antenna coupling, see the original paper for details). Dotted
and dashed lines are theoretical results for GOE and POE,
respectively, taking absorption into account. The dash-dotted line
in (b) corresponds to the theoretical expectation for the $C_3$
billiard (for details see the original work)
(taken from reference \cite{sch03a}).}
\end{figure}

As an example figure~\ref{fig:corrhole} shows the cross
correlation for a rectangular and the chaotic cavity, obtained
from reflection measurements at two different antennas
\cite{sch03a}. There was a strong absorption, which had to be
considered in the theory. Whereas for integrable systems an
exponential decay of the correlation function is expected, caused
by the absorption, for chaotic systems the correlation is always
close to zero, thus exhibiting the correlation whole. In both
cases the experimental results follow the predictions.
Discrepancies are found only for the rectangular billiard for
small times. They are caused by the presence of the antennas
making the systems pseudo-integrable and producing a small
correlation hole which is absent in the ideal rectangle
\cite{haa91b}.

An experimental observation of the correlation hole in the regime
of visible light has been achieved by Dingjan \etal \cite{din02}.
The authors studied a transmission spectrum of a He-Ne laser
through an open resonator made up of two end mirrors, either
curved or flat and a curved folding mirror. By misaligning the
latter one very large aberrations could be introduced thus making
the system chaotic. Taking the Fourier transform of the spectral
autocorrelation functions both for the regular and the disturbed
system, curves were obtained very similar to the ones shown in
figure~\ref{fig:corrhole} for the microwave system.

Very recently the study of scattering matrix correlation functions
has been extended to the subject of fidelity. The fidelity
amplitude is defined as the overlap integral of some initial state
with itself after the evolution under the influence of two
slightly different time evolution operators $U(t)$, $U'(t)$,

\begin{equation}
\label{eq:fid01}
f(t) = \left< \psi(0) \right| U^\dagger(t) \, U^{\prime}(t)
   \left| \psi(0)\right>
\end{equation}
The modulus square of the fidelity amplitude, the fidelity
$F(t)=|f(t)|^2$, had been proposed years ago by Peres to quantify
the stability of a quantum-mechanical system against perturbations
\cite{per84}. The renewed interest in the topic results from its
relevance for quantum-information processing. The fidelity
amplitude as defined in equation~(\ref{eq:fid01}), is
experimentally hardly accessible, as it requires  the knowledge of
the complete wave function. This was the motivation by Sch\"afer
\etal \cite{Sch05b}, to introduce a scattering fidelity, defined in
terms of scattering matrix elements of the perturbed and
unperturbed system.
In the weak coupling limit the scattering fidelity approaches the
ordinary fidelity (\ref{eq:fid01}).
Perturbation theory predicts an
exponential decay of the fidelity for $t<t_H$, and a cross-over to
a Gaussian decay for $t>t_H$. $t_H=\hbar/\Delta E$ is the
Heisenberg time, where $\Delta E$ is the mean level spacing
\cite{cer02}.

\begin{figure}
\includegraphics[width=\textwidth]{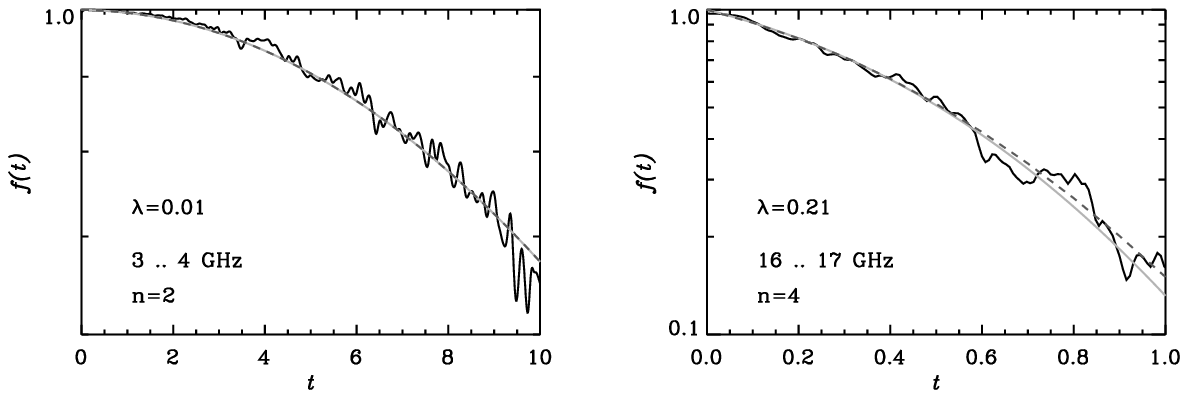}
\caption{\label{fig:fidelity} Fidelity amplitude for a chaotic
microwave billiard. The smooth solid line shows the
linear-response result, while the dashed line corresponds to the exact
result. The perturbation parameter $\lambda_{\rm exp}$
has been fitted to each experimental curve (see the original
work for details)
(taken from reference \cite{Sch05b}).}
\end{figure}

In a microwave experiment the perturbation was achieved by moving
one wall in a chaotic microwave resonator with two attached
antennas. In addition averaging was performed by superimposing the
results from different energy window and geometries. Figure~\ref{fig:fidelity}
shows the result for different perturbations.
For small perturbations the Gaussian decay dominates, whereas with
increasing perturbations strength a cross-over to an exponential
decay is observed. In the range accessible to the experiment a
good agreement was found with a linear-response prediction of
random matrix theory \cite{gor04}. Deviations of the linear
response results from the exact results, obtained by
super-symmetry techniques \cite{stoe04b,stoe05} amount to at most
10 percent in the regimes shown in the figure, which could not be
detected in the experiment.

Essentially the same results have been obtained in an experiment
in elastodynamic billiards \cite{lob03b}. They will be presented
in section \ref{subsec:SoundFidelity}.

\subsection{Current correlations}

A systems has to be opened if a measurement is to be performed.
As a consequence the eigenvalues are turned into resonances and
acquire widths.
This has been investigated in the previous subsections.
Now we want to investigate the effect of the openings on wave
functions and currents. Due to the openness the wave function
acquires an imaginary part, i.\,e.,
\begin{equation}\label{eq:PsiOpen}
\psi=\psi_r + \rmi \psi_i.
\end{equation}
A convenient parameter to characterise the openness is the phase
rigidity $|\rho|^2$ of $\psi$ \cite{lan97}, where
\begin{equation}\label{eq:RigidityDef}
 \rho = \int \rmd\bi{r}\,\psi(\bi{r})^2 =
 \frac{\langle\psi_r^2\rangle - \langle\psi_i^2\rangle}{\langle\psi_r^2\rangle +
 \langle\psi_i^2\rangle}\,.
\end{equation}
$\rho$=1 corresponds to the closed and $\rho$=0 to the completely
open system.

\begin{figure}
\includegraphics[width=\columnwidth]{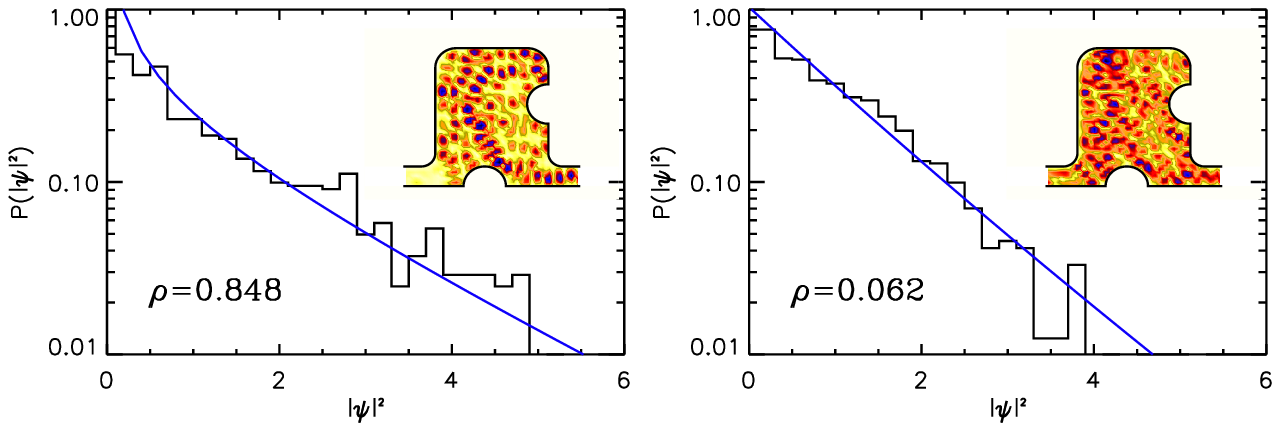}
\caption{\label{fig:PRhoI} Intensity distribution $P_\rho(I)$
for two different values of rigidity $\rho$. The insets show the
corresponding intensity pattern $I=|\psi|^2$ for a quantum dot
like microwave billiard.}
\end{figure}

The random plane wave approach introduced in
section~\ref{subsec:RPWA} allows to calculate the intensity
distribution. In open systems the amplitudes $a(\bi{k})$ (see
equation~(\ref{eq:2.20})) obey the relation $\langle a(\bi{k})
a(-\bi{k}) \rangle =  \rho \langle a(\bi{k})
a^*(\bi{k}) \rangle$. For a fixed value of $\rho$ this implies
a generalised Porter-Thomas distribution for the intensity,
\begin{equation}\label{eq:PRhoI}
  P_{\rho}(I)= \frac{1}{\sqrt{1 - |\rho|^2}}
  \exp \left[- \frac{I}{1 - |\rho|^2} \right]
  I_0 \left[ \frac{|\rho|I}{1 - |\rho|^2} \right],
\end{equation}
In figure~\ref{fig:PRhoI} the intensity distribution for a microwave
billiard with attached leads is plotted. The transitional
behaviour from a closed system to an open system is clearly
visible \cite{kim05a}. Good agreement with  the theoretical
prediction is found, where the only free parameter $\rho$ is taken
directly from the data.

\begin{figure}
\includegraphics[width=\columnwidth]{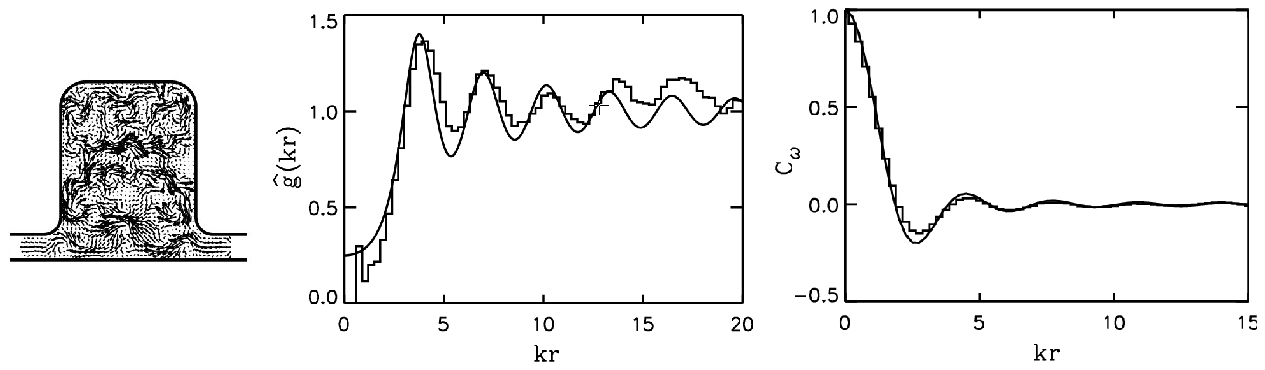}
\caption{\label{fig:FlowCwCVP}
 Left: Typical probability current pattern.
Centre: Experimental vortex pair correlation function $\hat{g}$,
together with the theoretical prediction. Right: vorticity spatial
autocorrelation function (taken from reference \cite{kim03a}).}
\end{figure}

For closed systems there is a large interest in nodal lines (see
e.\,g.~\cite{ber02b}), whereas in open systems there exists only
nodal points. The nodal points correspond to vortices of the
quantum mechanical probability density
\begin{equation}\label{eq:current}
  \bi{j}(\bi{r}) = \frac{A}{k}{\rm
  Im}[\psi^*(\bi{r})\nabla\psi(\bi{r})].
\end{equation}
At the same time the vortices give rise to phase singularities
which have been demonstrated in a rectangular microwave billiard
with an introduced absorber \cite{seb99}. There are a number of
theoretical  results on the distribution of the modulus of the
currents $j$ or its components $j_x$, $j_y$ \cite{sai02b}, vortex
spacing distributions \cite{sai01,ber02a}, various spatial
autocorrelation functions etc., using the plane wave
approximation. In quasi-tow-dimensional systems there is a
one-to-one correspondence between the quantum probability current
and the Poynting vector \cite{seb99}, allowing to determine the
flow patterns experimentally. The left part of
figure~\ref{fig:FlowCwCVP} shows a typical example. This has been
used to verify these distributions and correlation functions by
microwave experiments \cite{bar02,kim03a}. In the central
figure~\ref{fig:FlowCwCVP} the vortex pair correlation function is
shown, together with the prediction from the random plane wave
approximation \cite{ber00b,sai01}. In the right part of the figure
the spatial autocorrelation function of the vorticity $\omega=
(1/2)\bi{[\nabla\times j]}$ is shown, again together with the
theoretical prediction. In both examples a good agreement between
theory and experiment is found.

Usually spatial correlations vanish for $kr\to\infty$. Therefore
it came as a surprise when Brouwer showed that there exist a
number of correlation functions with long-range correlations
\cite{bro03}, i.\,e., with a surviving correlation in the limit
$kr\to\infty$. This is the case, e.\,g. for the connected correlator
of the squared intensity, $\langle I(\bi{r})^2I(\bi{r}')^2
\rangle_c$
showing a surviving correlation of $\mathrm{var} |\rho|^2$ for
$kr\to\infty$ (the subscript ``$c$'' refers to the connected
correlator, $\langle A B \rangle_c = \langle A B \rangle - \langle
A \rangle \langle B \rangle$). The long-range correlation was
experimentally verified for various correlators \cite{kim05a}.

\subsection{Classical phase space signatures of scattering}
\label{subsec:class}

In the preceding section we discussed some universal properties of
the scattering matrix. In the following a number of features shall
be presented associated with individual system properties. The
basis for the semi-classical treatment of scattering is an
expression for the scattering matrix originally given by Miller
\cite{mil75b},

\begin{equation}\label{class01}
    S(\bi{r},\bi{r}\,',k)=\sum\limits_n
    w_ne^{\frac{\imath}{\hbar}S_n}\,,
\end{equation}
where the sum is over all classical trajectories from $\bi{r}$ to
$\bi{r}\,'$. $w_n$ is a stability factor, which can be calculated
from the classical dynamics, and $S_n$ is the classical action. For
billiard system $S_n$ is given by $S_n=\hbar kl_n$, where $l_n$ is
the length of the trajectory. Equation (\ref{class01}) can be looked
upon as a special result of Gutzwiller's semi-classical quantum
mechanics, expressing the Green's function in terms of classical
trajectories \cite{gut90}.

In billiards systems the Fourier transform of equation
(\ref{class01}), with $k$ as the variable, gives the stability
weighted length spectrum. This has be used by Kim \etal
\cite{kim02,kim03a} in a number of papers to test a conjecture
that the transport in quantum dots is mainly promoted via scarred
wave functions. In a soft-walled microwave billiard the same
authors found evidence for dynamical tunneling, i.\,e.
quantum-mechanical cross-talking between stable islands in phase
space being classically separated \cite{kim05b}.

For a large number of open channels the scattering matrix in the
time domain decays exponentially with a time constant $\tau$,
resulting in a Lorentzian shape of the corresponding
autocorrelation function with width $\Gamma=1/\tau$, see
equation~(\ref{eq:2.47}). Semiclassical theory identifies $\tau$ with
the classical escape rate \cite{lew91,dor91}. This was the issue
of the first microwave experiment on scattering in an elbow-shape
resonator, where, however, the number of open channels was not
large but just one. Lu \etal \cite{lu99} studied the spectral
autocorrelation function in an $n$-disc scattering system with
$n=1,2,3,$ and found a perfect agreement between the observed
width and the values expected from the classical escape weight.

A characteristic feature of the $n$-disc system is a classical
repeller associated with a bouncing ball orbit between two
neighbouring discs. It follows from semiclassical theory that the
decay of the scattering matrix is no longer single-exponential but
given by a superposition,
\begin{equation}\label{class02}
    \hat{C}_{ij}(t)
    =\sum\limits_n a_n \rme^{-\gamma_n t}\cos({\gamma_n}' t)\,.
\end{equation}
where ${\gamma_n}'$ and $\gamma_n$ are real and imaginary part of
the eigenvalues of the Frobenius-Perron operator describing the
classical dynamics of the system. The corresponding autocorrelation
function shows the so-called Ruelle-Pollicott resonances.

These resonances have been studied by Pance \etal \cite{pan00a}
for the $n$-disk system for $n=1,2,3$. In all cases the found
resonances were in perfect agreement with those expected from the
classical dynamics.

Another scattering experiment showing clear fingerprints of the
underlying classical dynamics has been performed by Dembowski \etal
\cite{dem04a}. The authors studied the transport of microwaves
through a channel confined from above by a Gaussian and from below
by an inverted parabola. The dominant structure of the associated
Poincar\`e section, taken at either boundary, is a stable island
associated with the bouncing ball orbit along the symmetry line of
the channel. This region is classically not accessible. The
trajectories surround instead the stable island repeatedly in
phase space, before they leave the channel either to the left or
the right side. This revolution show up in a lighthouse-like
oscillatory decay of the scattering matrix on the time domain.

\subsection{Summary}

Up to about 1990 the study scattering was essentially a domain of theory.
Noticeable exceptions were the spectra of compound nuclei
giving rise to the development of random matrix theory \cite{por65}.
The situation changed completely with the emergence of different types
of billiard experiments. After the first microwave study \cite{stoe90}
there were numerous experiments with classical waves and quantum dots
\cite{stoe99}.
In microwave experiments, in contrast to quantum dots,
there is a complete control over the geometric parameters.
Therefore a quantitative agreement
between experiment and random matrix predictions was achieved
for all cases studied up to now. To this end it was necessary,
however, to take absorption and imperfect coupling into account
which was ignored in most previous investigations. Fortunately
there has been quite a large number of theoretical studies in
recent years to close this gap (\cite{fyoa}, this volume),
motivated last but not least by the possibility to test the
theoretical predictions experimentally.

\section{Experiments with sound waves}
\label{sec:SoundWaves}

\subsection{Measurements of eigenstatistics in acoustics}
\label{subsec:SoundEigenstatistics}

Manfred Schr\"oder \cite{sch54} appears to have been the first to
note that RMT level statistics should apply to acoustic systems
and could have implications for room acoustics and for structural
vibration.
Interestingly, in his thesis he used microwaves in
order to conduct experiments that would be analogous to an
acoustic system. Weaver \cite{wea89double,del94} was the first to
measure RMT statistics in acoustics. He determined about 150
elasto-dynamic levels in each of three small aluminum blocks
excited by impulsive forces and measured in the time domain with
piezoelectric detectors. Acoustic systems often lend themselves
naturally to time-domain measurements. Such measurement are
related by Fourier transform to the frequency sweeps typically
done on microwave billiards. Ellegaard and several co-workers
repeated these measurements, and extended them \cite{ell95,and01}.
Here Ellegaard \etal measured $\sim$~300 resonances in each of
several aluminum blocks with varying spherical octants of material
removed from a corner. Sub-Poissonian level statistics were
observed in the less irregular samples and ascribed to mode
conversions. Statistics approached the GOE as more material was
removed. The work was extended to quartz blocks \cite{ell96}, of
higher $Q$, with the identification of about 1400 levels. Again
various sized octants of material were removed from the corner,
thus allowing a study of the transition between a sub-Poissonian
and a GOE case. Neicu and Kudrolli \cite{nei02} also studied the
evolution of eigenstatistics under shape deformations.

Later work from the Copenhagen group emphasized studies of 2-d
plate-structures, analogous to the microwave billiard experiments
conducted by many others. At sufficiently low frequency or
small thickness of the plate it has only flexural and in-plane modes, which may
or may not be coupled depending on the breaking of up/down
symmetry. Bertelsen \etal \cite{ber00c} studied nearly 1000 levels
of a Sinai-stadium plate and found good agreement with a Weyl-like
estimate of level density, and, if up/down symmetry is broken,
good agreement with RMT predictions for other level statistics.
Andersen \etal \cite{and01} showed how variations in air-pressure
on such plates allows to distinguish between flexural waves (with
non-negligible losses to air) and in-plane modes (with low losses
to air). Once distinguished in this manner, the level statistics of
the two kinds of modes could be separately shown to be GOE.

Level dynamics, under variation of a parameter, has attracted
attention also. Not only can one study the evolution of statistics
as a system changes its symmetry or integrability, RMT also makes
predictions for the (scaled) distribution of level curvatures as a
parameter is varied. Bertelsen \etal \cite{ber99b} studied the
'Devil's spaghetti' formed by the level dynamics under variations
in temperature.

At a lower confidence level, due to lower $Q$ systems and few
modes identified, others also have confirmed GOE level statistics
in vibrations and acoustics, \cite{tei00,dav90}

The spatial statistics of diffuse fields in general and modes in
particular, have also attracted attention. There are a few
measurements directly imaging mode shapes, and a few of more
generic diffuse fields. More commonly such statistics are
investigated implicitly by means of field correlations amongst a
discrete set of points.

\subsection{Spatial correlations}
\label{subsec:SoundCorrelations}

Ebeling and Rollwage \etal \cite{rol85,ebe84} discussed
measurements of diffuse waves in reverberation rooms and in 2-d
ripple tanks and showed agreement with predictions based on the
random plane-wave superposition argument. In particular they
showed that field-field correlations
$\langle\psi(\bi{x})\psi(\bi{x}+\bi{r})\rangle =
J_{0}(k|r|)$, and that intensity-intensity correlations were the
expected $\langle I(\bi{x}) I(\bi{x}+\bi{r})\rangle =
2J_{0}^{2}(k|r|)+1$. The particular virtue of ripple tanks is
ready imaging of wavefields. Chinnery \cite{chi96} has also
presented pictures of (2-d) acoustic modes in water. Similar
studies have been made on drum heads \cite{tei00}, and capillary
waves \cite{blue92a}.

Schaadt \etal \cite{sch03b} studied mode shapes and their
statistics in a plate with flexural and in-plane modes. By
scanning their needle-like piezoelectric sensor, they could map a
component of the surface displacement field of any mode of their
choice. By means of pressure variations, they could determine if
the mode in question was flexural or in-plane (or a mixture if the
plate lacked up-down symmetry.) They showed the expected $1 + 2
J_{0}^{2}$ intensity-intensity ($I$-$I$) statistics amongst the
flexural waves, \textit{and} an expected more complex relation for
the $I$-$I$ statistics of the in-plane modes. The complexity is
due to the mixture of wave speeds associated with such modes
\cite{ako04a}.

Several works of late have reported theory and measurements of
diffuse-field-diffuse field correlations with features beyond the
well-known $J_{o}$ behaviour. These features are somewhat weaker
and more difficult to discern than the short-distance $J_{o}$, but
they are non-universal and of some practical importance. While
this work has generally been confined to correlations amongst
generic fields, one imagines that they apply also to modes.

\begin{figure}
\hspace{3cm}\includegraphics[width=8cm]{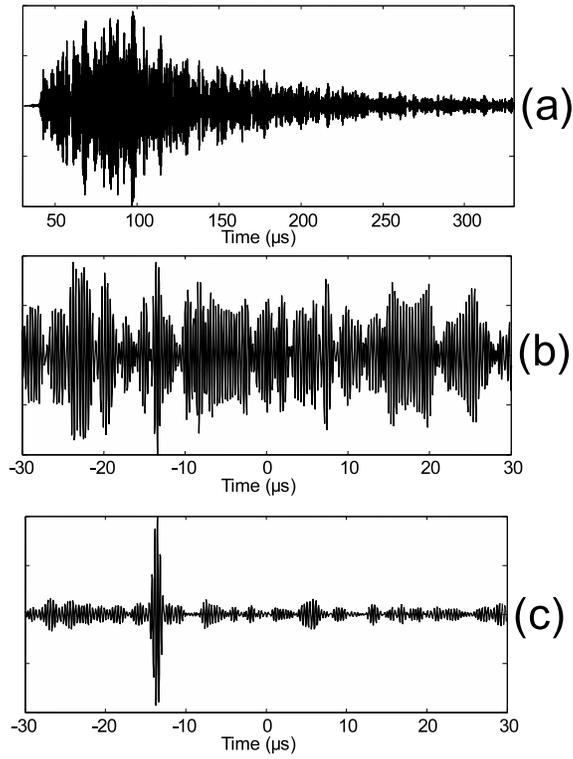}
\caption{\label{fig:weaver1}
A transient diffuse field in a multiply scattering water tank (a) is
autocorrelated (b). On averaging (b) over 118 distinct distant
sources, a sharp arrival appears. The sharp feature appears at a
time corresponding to a retroreflection from a nearby
heterogeneity, as would be seen if the receiver were used in
pulse-echo mode
(taken from reference \cite{lar04b}).}
\end{figure}

Weaver \cite{wea01a,wea01b} showed that the field-field
correlation function in a reverberant body was the
(imaginary part of the) Green's function. This is essentially
equivalent to a correlation between S matrix elements $\langle
S_{ij}(\omega) S_{kj}^{\ast}(\omega)\rangle$. The correlation
reduces to $J_{0}$ in the interior of a large 2-d single wave
speed isotropic billiard, but has other features in more complex
bodies, or near boundaries or scatterers. In particular it retains
signatures associated with reflections from nearby
heterogeneities. Weaver derived this from a modal perspective
\cite{wea01a,wea01b} using assumptions on modal statistics, but it
has also been derived in other ways. Derode \cite{der03} showed
that this relation is closely related to time-reversed acoustics.
An example, taken from \cite{lar04b} and plotted in figure~\ref{fig:weaver1},
shows how a conventional waveform, with sharp
arrivals in the time domain related to strong reflections, emerges
from correlations of a diffuse field. Campillo \etal
\cite{sha05,cam03,sab05} have used such correlations to map the
earth's surface structure using seismic waves. The ability to retrieve the Green's
function by passive measurements is gathering attention in ocean
acoustics and seismology \cite{sha05,cam03,sab05,rou05,rou03}.

\begin{figure}
\hspace*{3cm}\includegraphics[width=8cm]{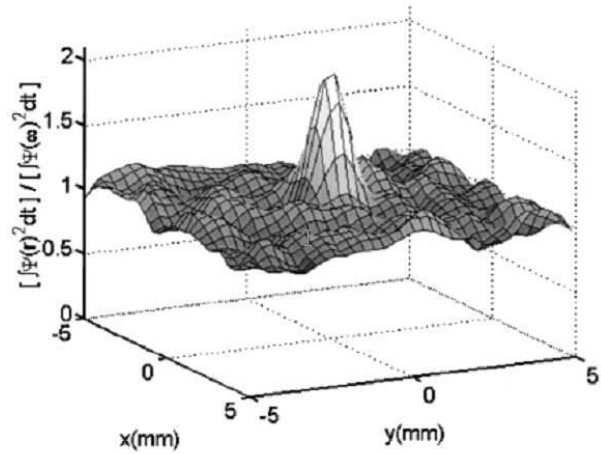}
\caption{\label{fig:weaver2}
The enhanced backscatter peak near
the position of the source in a 2-d chaotic elastic wave billiard
was imaged by de~Rosny \etal (taken from reference \cite{ros00}).}
\end{figure}

\subsection{Fidelity}
\label{subsec:SoundFidelity}

In section \ref{subsec:corr} the concept of scattering fidelity
was introduced to account for the fact that fidelity as originally
defined is not accessible to the experiment. It was shown that
under the assumptions of ergodicity the two definitions of
fidelity are the same \cite{pro02c,stoe04b}. Snieder introduced
the notion of `coda-wave interferometry' in seismology
\cite{sni02} equivalent to scattering fidelity. Weaver and Lobkis
independently studied the same idea in ultrasonics \cite{lob03b}
where temperature changes were responsible for the changed
Hamiltonian. De~Rosny \etal \cite{ros03} looked at a similar effect
with acoustic waves in a tank with moving scatterers. Snieder
applied it to ultrasonics in rock \cite{sni02} and showed an
effect from temperature variations and from the opening of a
crack. Weaver and Lobkis \cite{lob03b} showed that the measurement
could be used to evaluate the mixing rates of different waves. In
particular they showed that a regular body had much lower fidelity
than did a ray-chaotic body. They confined their quantitative
evaluations to times short compared to the Heisenberg time. Later
Gorin \etal \cite{gor2} showed that RMT theory for the time
dependence on the scale of the Heisenberg time was consistent with
Weaver and Lobkis's measurements over such times.

Ribay \etal have considered the effect of
temperature variations on reverberant acoustic signals in water
and in air and on flexural waves in plates \cite{rib05}. To the extent that
temperature changes in these systems merely change wave speed and
volume, and do so homogeneously, the only effect of such changes
is to dilate the waveforms. This trivial effect is not difficult
to account for.
Because longitudinal and shear wave speeds have different
dependencies on temperature, in elastic bodies there is also a
nontrivial change of waveform \cite{lob03b}.

Lu and Michaels \cite{lu05}
have suggested that ultrasonic fidelity, after correction for
temperature variations, may be useful as a nondestructive means
for monitoring the condition of materials and built structures.
Growth of a crack for example will degrade fidelity \cite{sni02},
and do so in a manner depending on the size and position of the
crack.

Lobkis (unpublished) has measured the almost perfect fidelity of
ultrasonic waveforms under the addition of a very weakly coupled
loss channel, and shown extraordinary sensitivity to that
addition, far greater than that exhibited by the energy dynamics
(see below)

\begin{figure}
\hspace*{3cm}\includegraphics[width=8cm]{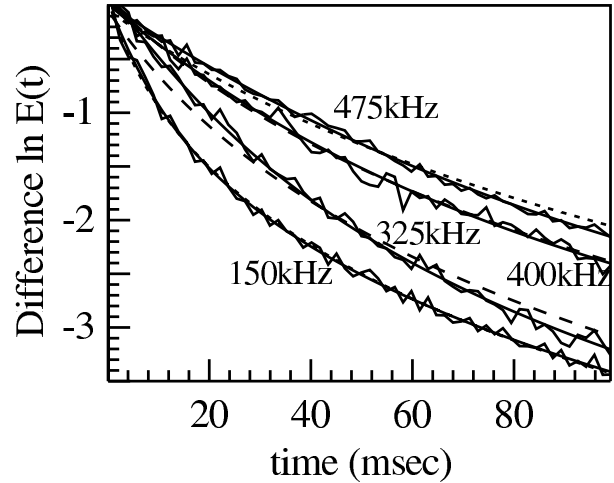}
\caption{\label{fig:weaver3}
Measurements of the free decay of
intensity $\vert S(t)\vert ^{2}$. It is found that the
Breit-Wigner formula (\ref{eq:weaver13}) and its generalizations
(dashed lines), often do not match measurements. A full RMT
calculation (solid lines) does match. Differences are most
pronounced when channels are well coupled
(taken from reference \cite{lob03a}).}
\end{figure}

\subsection{Reflection and Transmission Distributions}
\label{subsec:SoundTransmission}

Coherent backscattering, i.\,e., enhancements of the diagonal
elements of $G$ or $S$, are most profitably studied in the time
domain. Acoustics is thus a natural way to examine it. Coherent
backscatter enhancements by factors varying in time from two and
three are readily observed: $\vert S_{ii}\vert ^{2} \sim (2$  to
$3)\times \vert S_{ij}\vert^{2}$ $(i \ne j)$. Lyon \cite{lyo69} was
the first to predict enhanced backscatter in acoustics, followed
much later by a prediction of its dynamics by Weaver and Burkhardt
\cite{wea94} using a modal perspective and assumed GOE modal
statistics. It was first observed in acoustics by de~Rosny \etal
in a 2-d billiard \cite{ros00} (See fig.~\ref{fig:weaver2}) and by
Lobkis and Weaver in a 3-d billiard \cite{wea00b}. Larose \etal
\cite{lar04a} observed enhanced backscatter with seismic waves in
an open multiply scattering system, with potential for application
to characterizing the earth.

A system consisting of two or more weakly coupled cavities can
exhibit an enhanced backscatter that is a form of Anderson
localization; all elements $S_{ij}$ or $G_{ij}$ with $i$ and $j$
within the same substructure are enhanced, by arbitrarily large
factors, relative to others. This was observed by Lobkis and
Weaver \cite{wea00a} in a system consisting of two coupled
reverberant elastic bodies. It has been analyzed theoretically by
Weaver and Lobkis \cite{wea00a} and by Gronqvist and Guhr
\cite{groe05}

In a non-lossy system, responses are given simply in terms of real
modes and real natural frequencies. Spectra $\vert
S_{ij}(\omega)\vert ^{2}$ consist of distinct resonances;
interpretation of spectra is straightforward. In the presence of
loss these resonances spread and at high enough loss, they
overlap, thus complicating interpretation of power spectra. At
very high loss where the modes are well overlapped, the statistics
simplify; they become equivalent to Ericson fluctuations
\cite{eri63}; responses in the time domain can be understood as
simply decaying coloured Gaussian random processes. Frequency
domain response statistics are derived from that picture by
Fourier transform.

Statistics at intermediate overlap are not as well understood.
Most acoustics has concentrated on the question of the
\textit{variance} of power transmission $\vert S_{ij}(\omega)\vert
^{2}$ for $i \ne j$, (as opposed to its \textit{distribution} as in
section \ref{subsec:PSPT}) These may be considered
intensity-intensity statistics, and thus an extension of the
field-field statistics of the previous section. Lyon \cite{lyo69}
early appreciated that these random spectra were relevant to
reverberation rooms and structural vibrations. He made theoretical
estimates for the intensity variance by expressing responses in
the modal form (\ref{eq:weaver7}), and taking expectations of
$\vert G\vert^{2}$ written in terms of sums of various products of
four modal factors. The square of the intensity $T=\vert
G\vert^{2}$ transmitted from a source at $y$ with polarization $j$
to a receiver at $x$ with polarization $i$ is given, after a few
minor approximations and simplifications, by the modal expansion
\cite{lob00a}:

\begin{equation}
\label{eq:weaver14}\fl T^2=\sum_{rmlk} \frac{u_i^r(\bi{x})
u_j^r(\bi{y})}{\omega-\omega_r+\rmi\Gamma_r}
\frac{\{u_i^m(\bi{x})
u_j^m(\bi{y})\}^*}{\omega-\omega_m+\rmi\Gamma_m}
\frac{u_i^l(\bi{x})
u_j^l(\bi{y})}{\omega-\omega_l+\rmi\Gamma_l}
\frac{\{u_i^k(\bi{x})
u_j^k(\bi{y})\}^*}{\omega-\omega_k+\rmi\Gamma_k}
\end{equation}

Under a naive assumption of simple eigenstatistics and
homogeneous damping, (real modes, fixed level width) and using
expressions like (\ref{eq:weaver14}), Lyon evaluated expectations
of $T$ and $T^{2}$. Waterhouse \cite{wat78} and Davy
\cite{dav81b,dav86,dav87} discussed and extended the analysis of
Lyon. Davy carried out extensive measurements in reverberation
rooms in order to corroborate such predictions, and found some
limited success. Weaver \cite{wea89b} pointed out that
consideration of spectral rigidity neglected by Lyon and Davy
would improve agreement, but that discrepancies would remain.
Lobkis \etal \cite{lob00a} re-developed the modal-expansion theory
for variance of $T$ using exact GOE level and mode statistics.
They also incorporated the effect of a level width distribution
and consequent decay curvature.
For lack of any other model, they
retained the assumption of real eigenmodes, with modal statistics
identical to those of the GOE.

\begin{figure}
\hspace*{3cm}\includegraphics[width=9cm,clip]{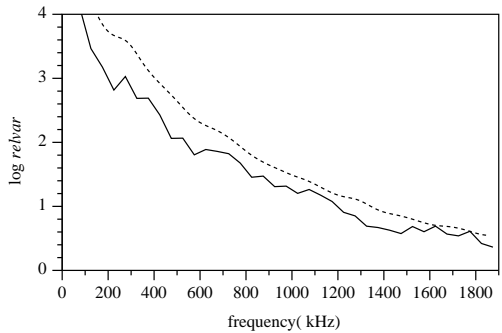}\\*[-7.5cm]
\caption{\label{fig:weaver4}
The normalized variance of power transmission $|G_{i j}|^2$ for $i \ne
j$, as measured (jagged line) in a reverberant aluminum block, and
as predicted (dashed line) using a `naive' theory like that of
Lyon \cite{lyo69} or Davy \cite{dav81b,dav86,dav87} but
incorporating GOE statistics on  eigenvalues and real
eigenfunctions, and a distribution of level widths. It is found
that the naive theory consistently overestimates the variance
(taken from reference \cite{lob00a}).}
\end{figure}

Like that of Davy, their theory consistently overestimated
variance relative to measurements on ultrasound in reverberant
elastic bodies, see figure~\ref{fig:weaver4}. They ascribed the
discrepancies to their neglect of the complexness of the
eigenmodes. A detailed comparison with a direct supersymmetric RMT
calculation for variance (e.\,g. \cite{roz04}) has not yet been
made.

Langley \etal \cite{cot04,lan04} have continued to study the
problem of variance in statistical structural vibration, and have
presented evidence for a consistent difference between the RMT
value for inverse participation ratio (three) and that observed in
their structures. They furthermore conjecture that this difference
may explain the differences between measured and predicted power
variances.

\subsection{Dissipation}
\label{subsec:SoundDissipation}

In dissipative systems, losses can be intrinsic to the material or
the structure, or be due to explicit open channels. The first
order effect of losses is to introduce a uniform exponential decay
in the time domain, equivalently to add an imaginary constant to
the frequency. It may be a nuisance in experiments, but the
effects are trivial when viewed that way, and the theoretical
questions it engenders are simple. All eigen-statistics are
preserved.

In practice losses are not perfectly diagonal and there is some
interesting mesoscopic physics. Dissipation is often distributed
nearly homogeneously, so deviations from the predictions of simple
theory are small; they are nevertheless observable. Three major
non-trivial topics have been addressed. The simplest is perhaps
that of `decay curvature' \cite{kaw86,bod87,bur96,bur97,bur98}
(discussed below) in
which a variance of level widths manifests as a non-exponential
decay of intensity in the time domain as in equation~\ref{eq:2.46}
and \ref{eq:weaver13}. In the frequency domain this manifests in
field-field correlations $\langle S_{ij}(\omega
)S_{ij}^*(\omega+\Omega)\rangle$ (see section \ref{subsec:SoundCorrelations}).
There are also studies of
transmission power variances $\langle\vert
S\vert^{4}\rangle-\langle\vert S\vert^{2}\rangle^{2}$
(see section \ref{subsec:SoundTransmission})
and of complex modes (see section \ref{subsec:SoundComplexEigenmodes}).

Decay curvature (\ref{eq:weaver13}) was derived in acoustics by
Schr\"oder \cite{sch65} and later by Burkhardt
\cite{bur96,bur97,bur98}; it is based on a Breit-Wigner expression
for level widths (\ref{eq:2.42}), (\ref{eq:weaver5})
(It was discussed for microwaves in eqn.~\ref{eq:2.16}).
Burkhardt
\cite{bur96,bur97,bur98} found good agreement between observed
decay curvatures and the formula (\ref{eq:weaver13}) and
furthermore found that $\nu$ correlated with the number of sites
of material damage in a reverberant body. Weaver and Lobkis
\cite{lob03a} and Lobkis Weaver and Rozhkov \cite{lob00a} also
found that the formula (\ref{eq:weaver13}) does a good
phenomenological job of fitting observed decay curvatures over
accessible dynamic ranges. This is may be surprising, as the
assumption of equally strong loss channels is surely incorrect.
It is furthermore unreasonable to suppose that real modal
statistics assumed in the derivation of (\ref{eq:2.46}) and
(\ref{eq:weaver13}) are correct. Thus Weaver and Lobkis
\cite{lob03a} went on to show that, carefully scrutinized, decay
curvature does \textit{not} fit to the formula
(\ref{eq:weaver13}). They did find that a full supersymmetric RMT
calculation \cite{lob03b} does fit the observations
(see figure~\ref{fig:weaver3}).

It is possible to understand the difference between the
Breit-Wigner picture of decay curvature, equation~(\ref{eq:weaver13})
and (\ref{eq:2.46}), and the exact RMT result by noting that the
actual modes of a lossy system are complex. Thus the conclusion
that level widths are chi-square random numbers given by a sum of
squares of $\nu$ equal-variance Gaussian numbers is incorrect.
That sum must be replaced by a sum of $\nu$ squares of the real
parts of the modes and $\nu$ squares of their imaginary
parts, cf equation~(\ref{eq:weaver5}). Thus the level width
distribution ought be more narrow in the presence of complex
modes, and the decay curvature ought be less pronounced.
RMT indeed predicts less decay curvature than does Breit-Wigner,
especially for very open channels; see for example figure~\ref{fig:weaver3}.
Microwave results on non-algebraic decay have already been
presented in section~\ref{subsec:tests}.

\subsection{Complex eigenmodes}
\label{subsec:SoundComplexEigenmodes}

\begin{figure}
\hspace*{3cm}\includegraphics[width=10cm]{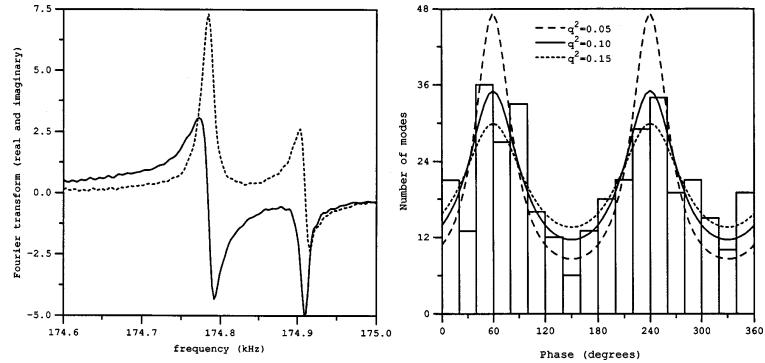}
\caption{\label{fig:weaver5}
A short section of the real and imaginary parts of a transmission through
a reverberant lossy elastodynamic body (left). It is clear that these two
modes differ by a substantial phase. A histogram of the phase
of resonances in this body fits well to a model of complex
Gaussian modes (right)
(taken from reference \cite{lob00b}).}
\end{figure}

Non-real modes, and in particular their statistics,
which have been discussed for microwaves in terms of rigidity
(see section~\ref{subsec:corr})
are at present under-studied in acoustics as well.
Lobkis and Weaver \cite{lob00b} showed that modes
in moderately damped structures (where they remain discernable as
they are only moderately overlapped) can have significant
complexity, and that their phases have a distribution in good
accordance with theoretical guesses (see figure~\ref{fig:weaver5}).
They found that observed complexities were sufficient to account
for differences like those of figure~\ref{fig:weaver4} between
naive theory and measurements. If theoretical estimates may be obtained
for complex modal statistics, perhaps in terms of level width
variances \cite{bar05b}, it may be that the modal-based estimates
for power variances
\cite{lyo69,dav81b,dav86,dav87,wea89b,lob00b,cot04,lan04} could be
brought into agreement with measurements.
The distribution of resonance phases shown in figure~\ref{fig:weaver5}b
is presumably related to that of the Poisson kernel
introduced for microwave systems in section~\ref{subsec:corr},
but the exact correspondence is not yet clear.

\subsection{Summary}

In acoustics the study of response statistics  has developed, over
decades, mostly within the context of eigenstatistics. Theoretical
predictions for field-field correlations and for intensity
correlations have been formulated in terms of assumptions about the
modes, with some success. It is only very recently that RMT theory
and better informed measurements have been applied and the various
assumptions questioned. For response statistics, as for decay
curvature, a direct RMT calculation, e.\,g., by supersymmetry can be
expected to match experimental measurements more perfectly than will
an eigenmode perspective. Nevertheless eigenmode perspectives
\cite{lyo69,dav81b,dav86,dav87,lob00b} that make convenient albeit
incorrect assumptions about eigenstatistics have had a good deal of
success. That approach to decay curvature is in most cases
sufficiently accurate, and is also adequate for field-field
correlations.

The eigenmode perspective for intensity variances has so far
failed to accurately match measurements. If, however, such
perspectives were to be extended beyond that of the most recent
\cite{lob00a}, one imagines that they might achieve an accuracy
approaching that of RMT. While a full RMT treatments of systems
with loss, using for example supersymmetric \cite{roz04} or
Monte-Carlo methods, are in principle possible, it remains
desirable also to have an approach in terms of the familiar and
useful and simpler concepts of eigenmodes. The estimates of
Lyon, Davy, Weaver, and Langley are close enough to success to
encourage the hope that such an approach may become feasible, and
that a modal picture can be developed to comprehend all aspects of
response statistics.

\section{Summary}

It is not unusual in science that related concepts are developed
independently in different fields without mutual knowledge of each
other. This is particularly true for scattering theory. The
example of Ericson fluctuations in nuclear physics on the one hand,
and universal conductance fluctuations in mesoscopic physics on
the other hand, and universal power transmission coefficient fluctuations
in acoustics, have already been mentioned. Another example is the
random plane wave approximation which was proposed
independently in the quantum mechanics of chaotic systems
\cite{ber77a} and in acoustics \cite{ebe84}.
Also the concept of fidelity has been independently
developed in both fields, though using different notions.
It has been the motivation for this review to illustrate the underlying concepts
common to all types of classical waves, with special emphasis on
microwaves and sound waves.

R.~W. was supported by the NFS and H.-J.~St. and U.~K. by the DFG.
\section*{References}

\bibliographystyle{rsiop}
\bibliography{thesis,paperdef,paper,newpaper,book,weaver}

\end{document}